\documentclass[prd,letterpaper,10pt,twoside,tightenlines,nofootinbib,showpacs,notitlepage]{revtex4-1}
\pdfoutput=1

\usepackage{amsmath,amsfonts,amssymb,latexsym}
\usepackage[sort&compress]{natbib}
\usepackage{grffile,epstopdf}
\usepackage{graphicx}
\usepackage{subcaption}
\usepackage{hhline}
\usepackage{xcolor}
\usepackage{hhline}

\newcommand{\m}{\cdot}

\newcommand{\res}{res}

\makeindex

\begin{document}
	
	\title{Diffusion processes involving multiple conserved charges: a first study from kinetic theory and implications to the fluid-dynamical modeling of heavy ion collisions}
	\author{Jan A. Fotakis}
	\email{fotakis@th.physik.uni-frankfurt.de}
	\affiliation{Institut f\"ur Theoretische Physik, Johann Wolfgang Goethe-Universit\"at,
		Max-von-Laue-Str.\ 1, D-60438 Frankfurt am Main, Germany}
	\author{Moritz Greif}
	\email{greif@th.physik.uni-frankfurt.de}
	\affiliation{Institut f\"ur Theoretische Physik, Johann Wolfgang Goethe-Universit\"at,
		Max-von-Laue-Str.\ 1, D-60438 Frankfurt am Main, Germany}
	\author{Gabriel Denicol}
	\affiliation{Instituto de F\'{i}sica, Universidade Federal Fluminense, UFF, Niter\'{o}i, 24210-346, RJ, Brazil}
	\author{Harri Niemi}
	\affiliation{Department of Physics, University of Jyv\"askyl\"a, P.O. Box 35, FI-40014 University of Jyv\"askyl\"a, Finland}
	\affiliation{Helsinki Institute of Physics, P.O. Box 64, FI-00014 University of Helsinki, Finland}
	\author{Carsten Greiner}
	\affiliation{Institut f\"ur Theoretische Physik, Johann Wolfgang Goethe-Universit\"at,
		Max-von-Laue-Str.\ 1, D-60438 Frankfurt am Main, Germany}
	\date{\today }
	
	\begin{abstract}
		The bulk nuclear matter produced in heavy ion collisions carries a multitude of conserved quantum numbers: electric charge, baryon number, and strangeness. Therefore, the diffusion processes associated to these conserved charges cannot occur independently and must be described in terms of a set of coupled diffusion equations. This physics is implemented by replacing the traditional diffusion coefficients for each conserved charge by a diffusion coefficient matrix, which quantifies the coupling between the conserved quantum numbers. The diagonal coefficients of this matrix are the usual charge diffusion coefficients, while the off-diagonal entries describe the diffusive coupling of the charge currents. In this paper, we show how to calculate this diffusion coefficient matrix from kinetic theory and provide results for a hadron resonance gas and a gas of partons. We further find that the off-diagonal entries can reach similar magnitudes compared to the diagonal entries. In order to provide some insight on the influence that the coupling between the net charge diffusion currents can have on heavy ion observables, we present first results for the diffusive evolution of a hadronic system in a simple (1+1)D-fluid dynamics approach, and study different configurations of the diffusion matrix.
	\end{abstract}
	
	\maketitle

	
	\section{Introduction}
	\label{sec:Intro}
	
	\nocite{Greif2017PRL} 
	
	The main motivation for studying nuclear collisions at relativistic energies is to understand the properties of strongly interacting matter. Especially the possibility of observing the transition from hadronic matter to quark-gluon plasma (QGP), as predicted by Quantum Chromodynamics (QCD), has been the driving force behind the active experimental heavy-ion programs at the Brookhaven National Laboratory (BNL) and the Conseil Europ\'{e}en Pour La Recherche Nucl\'{e}aire (CERN). During the last couple of decades, the high-energy nuclear collision experiments performed in the Relativistic Heavy Ion Collider (RHIC), at BNL, and in the Large Hadron Collider (LHC), at CERN, have shown that a considerable amount of QCD matter is produced in these collisions and that it is possible to infer the properties of such matter from the experimental data. 
	
	As a very prominent example several studies~\cite{Romatschke:2007mq,Xu:2007jv,Luzum:2008cw,Bozek:2009dw,Song:2010mg, Niemi:2011ix,Wesp:2011yy,Gale:2012rq,Niemi:2015qia,Bernhard:2016tnd} have demonstrated that QGP has one of the smallest shear viscosity to entropy density ratios in nature. Recently, much attention was given to the bulk viscosity of QCD~\cite{Arnold:2006fz,Kharzeev:2007wb,Bozek:2009dw,Noronha-Hostler:2013gga,Nopoush:2014pfa,Denicol:2014vaa,Ryu:2015vwa}, a coefficient which can display novel behavior near the deconfinement transition of nuclear matter. Moreover, several studies on the lattice \cite{Aarts:2014nba,Brandt:2015aqk,Ding:2016hua}, in perturbative QCD (pQCD) \cite{Greif:2014oia,Puglisi:2014sha,Arnold:2000dr}, and effective \cite{Rougemont:2015ona,Greif:2016skc,Rougemont:2017tlu} or kinetic theories \cite{Hammelmann:2018ath} have recently studied the electric conductivity. This coefficient is important in magnetohydrodynamical simulations, see e.g.~Ref.~\cite{Inghirami:2019mkc}.
	
	A dissipative process that is usually neglected, is the diffusion of conserved charges due to temperature or density gradients. Diffusion is a dissipative process which occurs as soon as inhomogeneities arise in a conserved quantity. In the simplest non-relativistic case, the diffusion current $\vec{j}_q$ of charge $q$ is described through Fick's first law \cite{Fick1855,Einstein1905},
	\begin{align}
		\vec{j}_q(t,\vec{x}) = -D_q \vec{\nabla} n_q(t,\vec{x}) \label{eq:FicksLaw},
	\end{align}
	where the current is generated by gradients in net charge density $n_q(t,\vec{x})$ and the diffusion coefficient $D_q$ characterizes the reaction strength of this thermal force. In the highest-energy nuclear collisions, the created matter has almost zero net baryon density at midrapidity, and the effects of diffusion are expected to be small in this region~\cite{Monnai:2012jc}. However, diffusion is expected to play an increasingly important role as the net baryon density increases with the decreasing collision energy.
	
	Recently, the Beam Energy Scan (BES) program was initiated at RHIC. In this program, nuclear collisions were systematically performed at lower energies in order to investigate the phase diagram and transport properties of nuclear matter at finite net baryon (and net electric charge) densities~\cite{Aggarwal:2010cw,Mohanty:2011nm,Mitchell:2012mx}. At beam energies down to, e.g., $\sqrt{s_{\mathrm{NN}}}=7.7~\mathrm{GeV}$ in the RHIC BES, the baryon chemical potential can reach values up to $\mu_{\mathrm{B}}\sim 400~\mathrm{MeV}$, which is significant when compared to the temperatures that are reached in this system~\cite{Odyniec:2013kna, Adamczyk:2017iwn}. Furthermore, the Facility for Antiproton and Ion Research (FAIR) at the Gesellschaft f\"ur Schwerionenforschung (GSI) in Darmstadt, Germany, and the Nuclotron-based Ion Collider facility (NICA) in Dubna, Russia, aim to generate and study compressed hadronic matter at large baryon densities~\cite{Friman:2011zz}. The theoretical description of those collisions could rely on diffusion dynamics. 
	
	The constituents of strongly interacting matter carry a multitude of conserved quantum numbers: baryon number, strangeness, electric charge, among others. As a result, the diffusion currents of the conserved charges must be coupled with each other. This multicomponent nature of diffusion in strongly interacting matter was first fully embraced in Ref.~\cite{Greif2017PRL}, where the full matrix of diffusion coefficients was computed, and it was subsequently found that the diffusion coefficients, describing the cross-coupling between the diffusion currents, are of the same magnitude as the ``normal'' (diagonal) diffusion coefficients. The purpose of this study is to complement Ref.~\cite{Greif2017PRL} and provide more details on the computation of the diffusion matrix for strongly interacting matter, as well as to provide an initial hydrodynamic calculation that illustrates the influence of the cross-couplings in relativistic nuclear collisions. As we will show, a novel phenomenon emerging from the coupling is a generation of regions of non-zero net strangeness from initially net strangeness neutral matter.
	
	This work is organized in two parts. In the first part we discuss the diffusion coefficients and in the second part we present a first investigation with fluid dynamics. In Section \ref{basic_definitions} we define the most important notations and expressions used in the paper. Section \ref{sec:charge_diffusion} provides a short review of diffusion in a relativistic gas with multiple conserved charges and introduces the diffusion coefficient matrix, which characterizes the coupling of the diffusion currents. We present the derivation of the diffusion coefficient matrix within a linear response approach from relativistic kinetic theory in Section \ref{sec:LinearResponse} and we further discuss its properties and results in relaxation time approximation (RTA) in Section \ref{sec:RTA}. The first part of this work is concluded with detailed discussions of the results for the coefficient matrix for a hadronic and a massless partonic system in Sections \ref{sec:HadronicGasResults} and \ref{sec:massless}. In Section \ref{sec:Hydro} we provide a short overview of the fluid dynamic approach used and also present our first results for the longitudinal diffusive evolution of a hadronic system. A summarizing conclusion and an outlook is provided in Section \ref{sec:Conclusion}. We use natural units, $\hbar=c=k_B=1$, and greek indices run from $0$ to $3$.

	
	\section{Foundations}
	\label{basic_definitions}
	
	\subsection{Basic definitions}
	
	Throughout this paper, we will express the momentum as $k^\mu$ and the coordinates as $x^\mu$. We denote the metric as $g^{\mu\nu}$ and impose the $(+,-,-,-)$-signature. It is convenient to express all tensors in terms of irreducible tensors regarding the local fluid velocity, $u^\mu \equiv u^\mu(x)$. Therefore, we introduce the orthogonal projectors $\Delta^{\mu}_{~\nu} \equiv g^\mu_{~\nu} - u^\mu u_\nu$ and $\Delta^{\mu\nu}_{\alpha\beta} \equiv \frac{1}{2}\left( \Delta^{\mu}_{~\alpha} \Delta^{\nu}_{~\beta} + \Delta^{\mu}_{~\beta} \Delta^{\nu}_{~\alpha} \right) - \frac{1}{3}\Delta^{\mu\nu}\Delta_{\alpha\beta}$. The projectors are symmetric ($\Delta^{\mu\nu} = \Delta^{\nu\mu} = \Delta^{(\mu\nu)}$ and $\Delta^{\mu\nu}_{\alpha\beta} = \Delta^{(\mu\nu)}_{(\alpha\beta)}$) and are orthogonal to the fluid velocity ($u^{\nu} \Delta^{\mu}_{~\nu} = 0$ and $u^\alpha \Delta^{\mu\nu}_{\alpha\beta} = 0$). More details can be found in Refs.~\cite{DeGroot,Molnar2016}. We denote the projected tensors as $A^{\langle \mu \rangle} \equiv \Delta^{\mu}_{~\alpha} A^{\alpha}$ and $A^{\langle \mu \nu \rangle} \equiv \Delta^{\mu \nu}_{\alpha\beta} A^{\alpha\beta}$. The four-derivative can then be decomposed into the comoving derivative $\mathcal{D} = u^\nu \partial_\nu$ and the projected derivative or gradient $\nabla_\mu = \Delta_\mu^{~\nu} \partial_\nu$: 
	\begin{align}
		\partial_\mu = u_\mu \mathcal{D} + \nabla_\mu.
	\end{align}
	For later use, we define the particle energy in the local rest frame (LRF) as $E_{i,\textbf{k}} \equiv u_\mu k^\mu_i$, $E_{i,\textbf{k}} \overset{\text{LRF}}{=} k^0_i = \sqrt{{\vec{k}}^2 + m_i^2}$, where the index $i$ refers to the particle's species. The state of the system is characterized by the single-particle distribution function of each particle species, $f_{i}(x,p)$. It can be decomposed into an equilibrium part, $f^{(0)}_{i,\mathbf{k}}$, and an off-equilibrium part, $\delta f_{i,\mathbf{k}}$, as $f_{i}(x,p) = f^{(0)}_{i,\mathbf{k}} + \delta f_{i,\mathbf{k}}$. We introduce the following notation for the integration measure:
	\begin{align}
	\mathrm{d}K_i \equiv \frac{\mathrm{d}^3 k_i}{(2\pi)^3 k^0_i} .
	\end{align}
	The momentum integrals over the distribution functions will be expressed using the following notation: 
	\begin{align}
	\Big\langle A^{\mu_1 \dots \mu_\ell} \Big\rangle_{i}&\equiv \int \mathrm{d}K_i A^{\mu_1 \dots \mu_\ell} f_{i,\mathbf{k}}, \\
	\Big\langle A^{\mu_1 \dots \mu_\ell} \Big\rangle_{i,0} &\equiv \int \mathrm{d}K_i A^{\mu_1 \dots \mu_\ell} f^{(0)}_{i,\mathbf{k}}, \\
	\Big\langle A^{\mu_1 \dots \mu_\ell} \Big\rangle_{i,\delta } &\equiv \int \mathrm{d}K_i
	A^{\mu_1 \dots \mu_\ell} \delta f_{i,\mathbf{k}}.
	\end{align}
	
	\subsection{Kinetic theory}
	
	The evolution of $f_{i}(x,k) =: f_{i,\mathbf{k}}$ is given by the Boltzmann equation,
	\begin{align}
		k_i^{\mu }\partial_\mu f_{i,\mathbf{k}} =\sum\limits_{j=1}^{N_{\text{species}}}C_{ij}(x,k), \label{eq:BoltzmannEq}
	\end{align}%
	where $C_{ij}$ is the collision term.	
	The energy-momentum tensor $T^{\mu\nu}$ and the net charge currents $N^\mu_q$ are expressed as the
	following momentum integrals of the single-particle distribution function \cite{DeGroot} 
	\begin{align}
		T^{\mu \nu } = \sum\limits_{i=1}^{N_{\text{species}}}\left\langle k^{\mu
		}k^{\nu }\right\rangle _{i}, \quad \quad N_{q}^{\mu } = \sum\limits_{i=1}^{N_{\text{%
					species}}}q_{i}\left\langle k^{\mu }\right\rangle _{i} \quad \text{with} \quad q \in \lbrace B,Q,S \rbrace,
	\end{align}%
	 and they fulfill the local conservation laws: $\partial_\nu T^{\mu\nu} = 0$ and $\partial_\mu N_{q}^\mu = 0$. It is convenient to decompose $T^{\mu \nu }$ and $N_{q}^{\mu }$ in terms of the fluid velocity field, $u^{\mu}$. Without loss of generality, we use Landau's definition of the fluid velocity \cite{landau1959course}, where $u^\mu$ is an eigenvector of $T^{\mu\nu}$ with an eigenvalue given by the energy density in the local rest frame of the fluid (LRF), $\epsilon$. That is, $T^{\mu \nu }u_{\nu }=\epsilon u^{\mu }$. The decompositions read
	\begin{eqnarray}
		T^{\mu \nu } =\epsilon u^{\mu }u^{\nu }-\Delta ^{\mu \nu }\left( P_{0}+\Pi
		\right) +\pi ^{\mu \nu }, \quad \quad  N_{q}^{\mu } = n_{q}u^{\mu }+j_{q}^{\mu }, \label{Nmuformel}
	\end{eqnarray}%
	where we introduced the local isotropic equilibrium pressure $P_0$ in the LRF, the bulk viscous pressure $\Pi $ in the LRF, the shear stress tensor $\pi ^{\mu \nu }$, the net charge densities $n_{q}$ with $q \in \lbrace B,Q,S \rbrace$ in the LRF, and the corresponding net charge
	diffusion currents $j_{q}^{\mu }$. The bulk viscous pressure, the shear-stress tensor and the diffusion currents represent the dissipative corrections in the energy-momentum tensor and the four-currents of the charges. The diffusion currents of the net charges $q$ are the main objects of our investigation and represent the charges diffusing orthogonally to the flow of the fluid.
	In this scheme, each introduced quantity can also be expressed as a contraction of the currents, $T^{\mu\nu}$ and $N^\mu_q$, with $u^{\mu }$ and $\Delta ^{\mu \nu}$, 
	\begin{eqnarray}
		\epsilon = u_{\mu }u_{\nu }T^{\mu \nu },\quad P_{0}+\Pi =-\frac{1}{3}\Delta _{\mu \nu }T^{\mu \nu }, \quad \pi ^{\mu \nu } = \Delta _{\alpha \beta }^{\mu \nu }T^{\alpha \beta }, \quad n_{q}=u_{\mu }N_{q}^{\mu }, \quad j_{q}^{\mu } = N_{q}^{\left\langle \mu\right\rangle}. \label{eq:DissCurr}
	\end{eqnarray}
	By specifying an equation of state, we can define the temperature and the chemical potentials for this system using the
	traditional matching conditions \cite{landau1959course},
	\begin{align}
		\epsilon =\epsilon_{\mathrm{eq}}(T,\mu_{B},\mu_{Q},\mu_{S}), \quad \quad n_{q} = n_{q,\mathrm{eq}}(T,\mu_{B},\mu_{Q},\mu_{S}) \quad \text{with} \quad q \in \lbrace B,Q,S \rbrace \label{eq:LandauMatch},
	\end{align}
	where $\epsilon_{\mathrm{eq}}$ and $n_{q,\mathrm{eq}}$ are the energy
	density and net charge densities of the system in local thermodynamic equilibrium, respectively. These quantities are calculated in kinetic theory by introducing the local equilibrium distribution function. In this work we want to restrict ourselves to classical statistics, and therefore the equilibrium distribution is given by the Maxwell-Juettner function
	\begin{align}
		f_{i,\mathbf{k}}^{(0)}=g_{i}\exp \left( -u_{\mu }k_i^{\mu }/T+\mu_i/T\right) , \label{eq:MaxwellJuettner}
	\end{align}
	where $\mu_{i}=B_{i}\mu _{B} + Q_{i}\mu _{Q} + S_{i}\mu _{S}$ is the chemical
	potential and $g_i$ is the degeneracy of the $i$-th species. Furthermore, the local equilibrium pressure is determined by the temperature and chemical potentials,
	\begin{align}
	P_0 \equiv P_0\left(T, \mu_{\mathrm{B}}, \mu_{\mathrm{Q}}, \mu_{\mathrm{S}} \right) .
	\end{align}

	
	\section{Net Charge Diffusion}
	\label{sec:charge_diffusion}
	In order to describe diffusion processes in relativistic fluids, a relativistic version of Fick's law must be employed. For a fluid with only one conserved charge, $q$, the relativistic Fick's law reads \cite{landau1959course,Eckart1940}:
	\begin{align}
	j_q^\mu = \kappa_q \nabla^\mu \left( \frac{\mu_q}{T} \right),
	\end{align} 
	where the diffusion current is generalized to be generated by a gradient in the corresponding thermal potential of the charge $\alpha_q \equiv \mu_q / T = \beta_0 \mu_q$, and $\beta_0 = 1/T$ is the inverse temperature. Note that (in flat Minkowski space) in the local rest frame $\nabla^\mu \equiv (0,-\vec{\nabla})$, and because of the sign, diffusion currents dissipate the existing inhomogeneities that originally generated the current. Often, instead of the charge diffusion coefficient, $\kappa_q$, the corresponding charge conductivity, $\sigma_q \equiv \kappa_q/T$, is used.
	We can relate $\kappa_q$ to $D_q$ (introduced in Eq.~\eqref{eq:FicksLaw}) by evaluating
	\begin{align}
	j^\mu_q = D_q \nabla^\mu n_q(\beta_0,\mu_q) = D_q \left( \frac{\partial n_q}{\partial \beta_0} \nabla^\mu \beta_0 + \frac{\partial n_q}{\partial \alpha_q} \nabla^\mu \alpha_q \right),
	\end{align}
	and imposing that the temperature is homogeneous, $\nabla^\mu \beta_0 = 0$, leading to
	\begin{align}
	j^\mu_q \overset{\beta_0 = \mathrm{const.}}{=} D_q \frac{\partial n_q}{\partial \alpha_q} \nabla^\mu \alpha_q \overset{!}{=} \kappa_q \nabla^\mu \alpha_q \quad \Rightarrow \quad \kappa_q = \frac{\partial n_q}{\partial \alpha_q} D_q .
	\end{align}
	As already stated, there are multiple conserved charges in nuclear matter: the baryon number, strangeness and electric charge. Moreover, the constituents of quark and hadronic matter carry multiple types of these charges, e.g. the proton carries baryon number and electric charge while the hyperons carry strangeness, baryon number and electric charge. Therefore, these constituents must react to multiple types of gradients in charge chemical potentials, in such a way that a gradient in baryon number does not only generate a baryon current, but can also produce currents in strangeness and electric charge (depending on the chemistry of the system). In order to account for this coupling, we introduced the \textit{diffusion coefficient matrix} in Ref.~\cite{Greif2017PRL}, which relates the charge diffusion currents to gradients in all thermal potentials, $\alpha_q$, as
	\begin{align}
	\begin{pmatrix}
	\begin{tabular}{c}
	$j^\mu_{\mathrm{B}}$ 
	\\ 
	$j^\mu_{\mathrm{Q}}$
	\\ 
	$j^\mu_{\mathrm{S}}$
	\end{tabular}%
	\end{pmatrix}&=%
	\begin{pmatrix}
	\begin{tabular}{ccc}
	$\kappa_{\mathrm{BB}}$ & $\kappa_{\mathrm{BQ}}$ & $\kappa_{\mathrm{BS}}$
	\\ 
	$\kappa_{\mathrm{QB}}$ & $\kappa_{\mathrm{QQ}} $ & $\kappa_{\mathrm{QS}}$
	\\ 
	$\kappa_{\mathrm{SB}}$ & $\kappa_{\mathrm{SQ}}$ & $\kappa_{\mathrm{SS}}$
	\end{tabular}%
	\end{pmatrix}
	\m
	\begin{pmatrix}
	\begin{tabular}{c}
	$\nabla^\mu \alpha_{\mathrm{B}} $
	\\ 
	$\nabla^\mu \alpha_{\mathrm{Q}} $
	\\ 
	$\nabla^\mu \alpha_{\mathrm{S}} $
	\end{tabular}
	\end{pmatrix}. \label{eq:NavierStokesTerms}
	\end{align}
	The objective of this paper is the evaluation and a first look at the possible dynamic implications of the complete diffusion matrix. In the first part of this work we will present a method of computation from relativistic kinetic theory.


	\section{Linear Response Theory: First-order Chapman-Enskog Expansion}
	\label{sec:LinearResponse}
	In this chapter, we present a method of evaluating the full diffusion coefficient matrix in relativistic kinetic theory. Here, we follow Refs.~\cite{Greif2017PRL, Greif:2016skc} and consider a dilute gas consisting of $N_{\text{species}}$ particle species, with the $i$-th particle species having degeneracy $g_{i}$, electric charge $Q_i$, strangeness $S_i$ and baryon number $B_i$. The system shall be under the influence of spatial gradients in baryon, strangeness and electric chemical potentials over temperature $\nabla^\mu (\mu_{q}/T)$ (with $q \in \lbrace B,Q,S \rbrace$), but no other external forces, as assumed in Ref.~\cite{Greif:2016skc}. The gradients are assumed to be small, such that the distortions from (local) equilibrium are small and linear response theory is applicable.
	
	\subsection{The linearized Boltzmann equation}
	
	We consider the system to be initially in global equilibrium. Next, we apply small gradients in the chemical potentials that are instantly switched on and cause a small perturbation $\delta f_p^i$ of the single-particle distribution function (of the $i$-th particle species) from equilibrium. This perturbation generates a diffusion current in the corresponding charges. The aim of this section is to set up the Boltzmann equation for this situation. 
	
	The magnitude of gradients can be quantified by introducing the so-called Knudsen numbers, $\mathrm{Kn}$, which are constructed as ratios of the characteristic microscopic and macroscopic length scales, $\mathrm{Kn} = \ell_{\mathrm{micro}}/\ell_{\mathrm{macro}}$. 	
	Thus, Knudsen numbers are small if the corresponding macroscopic length scales, $\ell_{\mathrm{macro}}$, are large in comparison to the microscopic length scales, $\ell_{\mathrm{micro}}$. The later is often taken to be the mean free-path of a particle in the gas, and $\ell_{\mathrm{macro}}$ is related to the gradients in the system. If the gradients that generate the perturbation of the single-particle distribution are small, it may be possible to expand the distribution in terms of the Knudsen number and truncate such an expansion at lower order:
	\begin{align}
	f_{i,\textbf{k}} = f^{(0)}_{i,\textbf{k}} + \delta f_{i,\textbf{k}} = f^{(0)}_{i,\textbf{k}} + f^{(1)}_{i,\textbf{k}} + \mathcal{O}\left(\mathrm{Kn}^2\right).
	\end{align} 
	This expansion is also referred to as Chapman-Enskog expansion \cite{chapman1970mathematical}. If the Knudsen number is sufficiently small and the series defined above converges, it is possible to neglect contributions that are of second order or higher and the perturbed single-particle distribution function can be approximated solely in terms of its first order terms, $\delta f_{i,\textbf{k}} \approx f^{(1)}_{i,\textbf{k}} \sim \mathcal{O}(\mathrm{Kn})$. Applying the Chapman-Enskog expansion to the Boltzmann equation \eqref{eq:BoltzmannEq} and only retaining the terms that are of first order in the Knudsen number leads to the following equation:
	\begin{align}
	k_i^\mu \partial_\mu f_{i,\textbf{k}}^{(0)} = \sum_{j=1}^{N_{\text{species}}}\mathcal{C}_{ij}^{(1)}[f_{i,\textbf{k}}], \label{eq:LinBoltzEq}
	\end{align}
	where we introduced the linearized collision term
	\begin{widetext}
		\begin{align}
		\sum_{j=1}^{N_{\text{species}}} \mathcal{C}_{ij}^{(1)}[f_{i,\textbf{k}}] \equiv \sum\limits_{j=1}^{N_{\text{species}}} \sum\limits_{a=1}^{N_{\text{species}}} \sum\limits_{b=1}^{N_{\text{species}}}& \gamma_{ij} \int_{\mathbb{R}^3} \mathrm{d}P_a \int_{\mathbb{R}^3} \mathrm{d}P^\prime_b \int_{\mathbb{R}^3} \mathrm{d}K^\prime_j \, (2\pi)^6 s\, \sigma_{ij \rightarrow ab}(s,\Omega) \delta^{(4)}\left(k_i + k^\prime_j - p_a - p^\prime_b\right) \nonumber \\ 
		&\times f^{(0)}_{i,\textbf{k}} f^{(0)}_{j,\textbf{k}^\prime} \tilde{f}^{(0)}_{a,\textbf{p}} \tilde{f}^{(0)}_{b,\textbf{p}^\prime} \left(\frac{f^{(1)}_{i,\textbf{k}}}{f^{(0)}_{i,\textbf{k}} \tilde{f}^{(0)}_{i,\textbf{k}}} + \frac{f^{(1)}_{j,\textbf{k}^\prime}}{f^{(0)}_{j,\textbf{k}^\prime} \tilde{f}^{(0)}_{j,\textbf{k}^\prime}} - \frac{f^{(1)}_{a,\textbf{p}}}{f^{(0)}_{a,\textbf{p}} \tilde{f}^{(0)}_{a,\textbf{p}}} - \frac{f^{(1)}_{b,\textbf{p}^\prime}}{f^{(0)}_{b,\textbf{p}^\prime} \tilde{f}^{(0)}_{b,\textbf{p}^\prime}}  \right), \label{eq:LinCollTermGeneral}
		\end{align}
	\end{widetext}
	and $\sigma_{ij \rightarrow ab}(s,\Omega)$ is the differential cross section for the binary interaction of incoming particles of species $i$ and $j$, with outgoing particles of species $a$ and $b$ (denoted as $ij \rightarrow ab$), at the center of mass collision energy $\sqrt{s}$ in a solid angle $\Omega$. Further, we introduced the symmetry factor $\gamma_{ij} = 1 - \frac{1}{2} \delta_{ij}$ and the notation $\tilde{f}^{(0)}_{i,\textbf{k}} = 1 - a f^{(0)}_{i,\textbf{k}}$, where $a = 1$ for fermions, $a = -1$ for bosons or $a = 0$ for classical particles. In this paper, we limit our discussion to classical statistics and therefore $\tilde{f}^{(0)}_{i,\textbf{k}} = 1$, and to binary \emph{elastic} processes, where only processes $ij \rightarrow ij$ are considered ($\sigma_{ij \rightarrow ij} \equiv \sigma_{ij}$).
	
	The achieved equation is typical for perturbation theory: the perturbed quantity on the right-hand side is determined by unperturbed quantities on the left-hand side of the equation. The left-hand side in the linearized Boltzmann equation \eqref{eq:LinBoltzEq} can be evaluated by first decomposing the four-derivative into comoving time derivative and projected derivative, and then substituting the comoving time derivatives of the primary fields $\epsilon$, $n_q$ (or $\beta_0 \equiv 1/T$ and $\alpha_q \equiv \mu_q/T$ correspondingly) and $u^\mu$ using the explicit local conservation laws from ideal fluid dynamics,
	\begin{align*}
		\mathcal{D} \epsilon = -(\epsilon + P_0)\theta, \quad \quad (\epsilon + P_0)\mathcal{D}u^\mu = \nabla^\mu P_0, \quad \quad \mathcal{D} n_{q} &= -n_{q} \theta .
	\end{align*}
Above, we introduced the expansion scalar $\theta \equiv \nabla_\mu u^\mu$. Further, using the Euler relation,
	\begin{align}
	s = \beta_0 \left( \epsilon + P_0 \right) - \sum_{q \in \lbrace \mathrm{B}, \mathrm{Q}, \mathrm{S} \rbrace} \alpha_q n_q ,
	\end{align}
	and the Gibbs-Duhem relation, in the form,
	\begin{align}
	\beta_0\nabla^\mu P_0 = -s \frac{\nabla^\mu \beta_0}{\beta_0} +  \sum_{q \in \lbrace \mathrm{B}, \mathrm{Q}, \mathrm{S} \rbrace} n_q\left( \nabla^\mu \alpha_q - \frac{\alpha_q}{\beta_0} \nabla^\mu \beta_0 \right),
	\end{align} 
	we find the following equivalent form to the momentum conservation equation:
	\begin{align}
	\mathcal{D}u^\mu = - \frac{\nabla^\mu \beta_0}{\beta_0} + \sum_{q \in \lbrace B,Q,S \rbrace} \frac{n_q}{(\epsilon + P_0) \beta_0}\nabla^\mu \alpha_q.
	\end{align}
	Following this procedure, we derive the following source term \cite{Denicol:2011fa} for a system with multiple conserved charges (terms related to the shear-stress tensor and bulk viscous pressure are omitted in the last line)
	\begin{align}
		\mathcal{S}(x,k_i) &\equiv k_i^\mu \partial_\mu f^{(0)}_{i,\mathbf{k}} \nonumber\\
		&= -f^{(0)}_{i,\mathbf{k}} \left[ E^2_{i,\textbf{k}} \mathcal{D}\beta_0 - E_{i,\textbf{k}} \mathcal{D}\alpha_i + \frac{1}{3} ( m_i^2 - E^2_{i,\textbf{k}} ) \beta_0 \theta + \sum_{q\in\lbrace B,Q,S \rbrace} k^{\langle \mu \rangle}_i \nabla_\mu \alpha_q \left( \frac{E_{i,\textbf{k}} n_q }{\epsilon + P_0} - q_i \right) + \beta_0 k_i^{\langle\mu} k_i^{\nu\rangle} \sigma_{\mu\nu} \right] \nonumber\\
		&\simeq -\sum_{q\in\lbrace B,Q,S \rbrace}f^{(0)}_{i,\mathbf{k}} k_{i}^{\langle\mu\rangle} \nabla_\mu \alpha_{q} \left(\frac{E_{i,\mathbf{k}} n_{q}}{\epsilon + P_0} - q_i\right), \label{eq:FinalSourceTerm}
	\end{align}
	where we defined the shear tensor $\sigma_{\mu\nu} \equiv \partial_{\langle \mu} u_{\nu \rangle}$. The diffusion current is then calculated as,
	\begin{align}
		j^\mu_{q} = \sum\limits_{i=1}^{N_{\text{species}}} q_i \int \mathrm{d}K_i \, k_i^{\langle \mu \rangle} f^{(1)}_{i,\mathbf{k}}. \label{eq:current}
	\end{align}

Thus, the correction $f^{(1)}_{i,\mathbf{k}}$ related to net-charge diffusion can be calculated from the following linear equation, by inverting the linearized collision term,
	\begin{align}
		&\sum_{j=1}^{N_{\text{species}}}\mathcal{C}_{ij}^{(1)}[f_{i,\textbf{k}}]  = \mathcal{S}(x,p_i) = -\sum_{q\in\lbrace B,Q,S \rbrace}f^{(0)}_{i,\mathbf{k}} k_{i}^{\langle\mu\rangle} \nabla_\mu \alpha_{q} \left(\frac{E_{i,\mathbf{k}} n_{q}}{\epsilon + P_0} - q_i\right). \label{eq:FinalLinBoltzEq}
	\end{align}
	The source term can be understood as a force term that generates the perturbation of the single-particle distribution due to the gradients in the thermal potentials, $\nabla^\mu \alpha_q$, which will eventually give rise to diffusion currents in the conserved charges $q$, according to Eq. \eqref{eq:current}.
		
	\subsection{Deriving the explicit expression of the diffusion matrix}
	
	In the following sections, we derive explicit relations that are required for the derivation of the diffusion coefficient matrix. Following Refs.~\cite{Greif2017PRL,Greif:2016skc}, we can approximate the solutions $f^{(1)}_{i,\textbf{k}}$ of the linearized Boltzmann equation \eqref{eq:FinalLinBoltzEq} by expanding the first order perturbations in powers of energy and truncating the power series at the truncation order $M$:
	\begin{align}
	f^{(1)}_{i,\textbf{k}} = \sum_{q \in \lbrace B,Q,S \rbrace} f^{(0)}_{i,\textbf{k}} k_i^{\langle \mu \rangle} \nabla_\mu \alpha_{q} \sum_{m=0}^M \lambda_{m,q}^{(i)} E_{i,\textbf{k}}^m . \label{eq:ExpansionAnsatz}	
	\end{align}
	Applying this to the linearized collision term \eqref{eq:LinCollTermGeneral} in the classical limit and for elastic scatterings only results in
	\begin{widetext}	
		\begin{align}
			\sum_{j=1}^{N_{\text{species}}} \mathcal{C}_{ij}^{(1)}[f_{i,\textbf{k}}] = \sum_{q\in\lbrace B,Q,S\rbrace} &\nabla_\mu \alpha_q \sum_{m=0}^M \sum\limits_{j=1}^{N_{\text{species}}} \gamma_{ij}  \int_{\mathbb{R}^3} \mathrm{d}P_i \int_{\mathbb{R}^3} \mathrm{d}P^\prime_j \int_{\mathbb{R}^3} \mathrm{d}K^\prime_j \, (2\pi)^6 s\, \sigma_{ij}(s,\Omega) \delta^{(4)}\left(k_i + k^\prime_j - p_i - p^\prime_j\right) \nonumber \\ 
			&\times f^{(0)}_{i,\textbf{k}} f^{(0)}_{j,\textbf{k}^\prime} \left( \lambda^{(i)}_{m,q} \, k^{\langle \mu \rangle}_{i} E_{i,\textbf{k}}^m + \lambda^{(j)}_{m,q} \, {k^\prime}^{\langle \mu \rangle}_{j} E_{j,\textbf{k}^\prime}^m - \lambda^{(i)}_{m,q} \, p^{\langle \mu \rangle}_{i} E_{i,\textbf{p}}^m - \lambda^{(j)}_{m,q} \, {p^\prime}^{\langle \mu \rangle}_{j} E_{j,\textbf{p}^\prime}^m \right). \label{eq:ExpandedLinColl}
		\end{align}
	\end{widetext}
	We can rewrite the linearized Boltzmann equation into a matrix equation by multiplying Eq. \eqref{eq:FinalLinBoltzEq} with $E^{n-1}_{i,\textbf{k}} k^{\langle\nu\rangle}_i$ and then integrating over the momentum $k_i$, such that we evaluate orthogonal moments of the Boltzmann equation. This becomes even more apparent when we realize that $k_i^{\langle \mu \rangle}$ fulfills the following orthogonality relation
	\begin{align}
		\int \mathrm{d}K_i \, k_i^{\langle \mu \rangle} k_i^{\langle \nu \rangle} \mathcal{F} = \frac{\Delta^{\mu\nu}}{3} \int \mathrm{d}K_i \, k^{\langle \alpha \rangle}_i k_{i,\,\langle \alpha \rangle} \mathcal{F}, \label{eq:Orthogonality}
	\end{align}
	for arbitrary scalar functions $\mathcal{F}$ in energy. Furthermore, because the gradients in the thermal potentials, $\nabla^\mu \alpha_q$, are arbitrary, we can split the linearized Boltzmann equation into separate equations for each charge $q$. Using the above-mentioned evaluation of moments, the orthogonality properties of the momentum basis \eqref{eq:Orthogonality}, and the separation by charge, we arrive at the following set of linear equations \cite{Greif2017PRL}:
	\begin{align}
	\sum_{m=0}^M \sum_{j=1}^{N_{\text{species}}} \left( \mathcal{A}^{i}_{nm} \delta^{ij} + \mathcal{C}^{ij}_{nm} \right) \lambda^{(j)}_{m,q} = b^i_{q,n}, \label{eq:FinalSetLinEq}
	\end{align}
	where we introduced the abbreviations
	\begin{widetext}
		\begin{align}
			\mathcal{A}^i_{nm} &\equiv \sum_{\ell=1}^{N_{\text{species}}} \gamma_{i\ell} \int \mathrm{d}K_i \mathrm{d}K^\prime_\ell \mathrm{d}P_i \mathrm{d}P^\prime_\ell \, (2\pi)^6 s \sigma_{i\ell}(s,\Omega) \delta^{(4)}\left(k_i + k^\prime_\ell - p_i - p^\prime_\ell\right) f^{(0)}_{i,\textbf{k}} f^{(0)}_{\ell,\textbf{k}^\prime} E_{i,\textbf{k}}^{n-1} k_{i,\,\langle \alpha \rangle} \left( E^m_{i,\textbf{k}} k^{\langle \alpha \rangle}_i - E^m_{i,\textbf{p}} p^{\langle \alpha \rangle}_i \right), \nonumber \\
			\mathcal{C}^{ij}_{nm} &\equiv \gamma_{ij} \int \mathrm{d}K_i \mathrm{d}K^\prime_j \mathrm{d}P_i \mathrm{d}P^\prime_j \, (2\pi)^6 s \sigma_{ij}(s,\Omega) \delta^{(4)}\left(k_i + k^\prime_j - p_i - p^\prime_j\right) f^{(0)}_{i,\textbf{k}} f^{(0)}_{j,\textbf{k}^\prime} E_{i,\textbf{k}}^{n-1} k_{i,\,\langle \alpha \rangle} \left( E^m_{j,\textbf{k}^\prime} {k^\prime}^{\langle \alpha \rangle}_j - E^m_{j,\textbf{p}^\prime} {p^\prime}^{\langle \alpha \rangle}_j \right), \nonumber\\
			b^i_{q,n} &\equiv \int \mathrm{d}K_i \, E_{i,\textbf{k}}^{n-1} \left(m_i^2 - E_{i,\textbf{k}}^2 \right) \left( \frac{E_{i,\textbf{k}} n_q}{\epsilon + P_0} - q_i \right) f^{(0)}_{i,\textbf{k}},
		\end{align}
	\end{widetext}
	and we used the dispersion relation $k_{i,\,\langle \alpha \rangle} k_i^{\langle \alpha \rangle} = \Delta_{\alpha\beta} k^\alpha_i k^\beta_i =  m_i^2 - E_{i,\textbf{k}}^2$. Equation \eqref{eq:FinalSetLinEq} is an ordinary matrix equation, where $\mathcal{M}_{nm}^{ij} \equiv \mathcal{A}^i_{nm}\delta^{ij} + \mathcal{C}^{ij}_{nm}$ are the entries of an $\left[(N_{\text{species}} \cdot M) \times (N_{\text{species}} \cdot M)\right]$-matrix, $b^i_{q,n}$ are the entries of an $(N_{\text{species}} \cdot M)$-dimensional source vector and $\lambda^{(j)}_{q,m}$ are the entries of an $(N_{\text{species}} \cdot M)$-dimensional vector of the expansion coefficients from Eq. \eqref{eq:ExpansionAnsatz}, which are the solutions of the linear set of equations. In order to make matrix $\mathcal{M}$ quadratic, we set the parameter $n$ to run from 0 to $M$. Furthermore, there are as many sets of such matrix equations \eqref{eq:FinalSetLinEq} as there are considered charge types. In this paper, we limit ourselves to baryon number, strangeness and electric charge, and therefore there are three sets of linear equations to solve.
	
	Up until this point, all steps were done without imposing the definition of the local rest frame. As already stated, in this work, we use the Landau definition of the four-velocity \cite{landau1959course}, in which all orthogonal energy-momentum flow vanishes:
	\begin{align}
		W^\mu \equiv \sum_{i=1}^{N_\text{species}} \left\langle E_{i,\textbf{k}} k^{\langle \mu \rangle}_i \right\rangle_{i,\delta} \overset{!}{=} 0. \label{eq:EnergyFluxLandau}
	\end{align}
	Applying the expansion \eqref{eq:ExpansionAnsatz} of $f_{i,\textbf{k}}^{(1)}$ to Eq.~\eqref{eq:EnergyFluxLandau} gives us an additional constraint for the expansion coefficients,
	\begin{align}
		\sum_{i=1}^{N_{\text{species}}} \sum_{m=0}^{M} \lambda^{(i)}_{m,q} \left\langle E^{m+1}_{i,\textbf{k}} \left( m_i^2 - E_{i,\textbf{k}}^2 \right) \right\rangle_{i,0} = 0. \label{eq:Constraint}
	\end{align}
	Together with the matrix equations \eqref{eq:FinalSetLinEq}, Eq.~\eqref{eq:Constraint} forms a set of linear equations of which the expansion coefficients $\lambda^{(i)}_{q,m}$ are the solutions. By applying the expansion in Eq.~\eqref{eq:ExpansionAnsatz} to the diffusion current \eqref{eq:current} and directly comparing to its Navier-Stokes form,
	\begin{align}
		 \sum_{q^\prime \in \lbrace B,Q,S \rbrace} \kappa_{qq^\prime} \nabla^\mu \alpha_{q^\prime} \overset{\text{Navier-Stokes}}{=} j^\mu_q &\equiv \sum_{q^\prime \in \lbrace B,Q,S \rbrace} \nabla_\nu \alpha_{q^\prime} \sum\limits_{i=1}^{N_{\text{species}}} \sum_{m=0}^M \lambda_{m,q^\prime}^{(i)} q_i \int \mathrm{d}K_i \, E_{i,\textbf{k}}^m k_i^{\langle \mu \rangle} k_i^{\langle \nu \rangle} f^{(0)}_{i,\textbf{k}} \nonumber\\
		&= \frac{1}{3} \sum_{q^\prime \in \lbrace B,Q,S \rbrace} \nabla^\mu \alpha_{q^\prime} \sum\limits_{i=1}^{N_{\text{species}}} \sum_{m=0}^M \lambda_{m,q^\prime}^{(i)} q_i \int \mathrm{d}K_i \, E_{i,\textbf{k}}^m \left( m_i^2 - E_{i,\textbf{k}}^2 \right) f^{(0)}_{i,\textbf{k}} ,
	\end{align}
	we arrive at an explicit form for the entries of the diffusion coefficient matrix:
	\begin{align}
		\kappa_{qq^\prime} = \frac{1}{3} \sum\limits_{i=1}^{N_{\text{species}}} q_i \sum_{m=0}^M \lambda_{m,q^\prime}^{(i)} \int \mathrm{d}K_i \, E_{i,\textbf{k}}^m \left( m_i^2 - E^2_{i,\textbf{k}} \right) f^{(0)}_{i,\textbf{k}} . \label{eq:FinalLinearizedDiffu}
	\end{align}
	In Ref.~\cite{Greif:2016skc}, the quick convergence of the series in Eq.~\eqref{eq:FinalLinearizedDiffu} was demonstrated and thus, we restrict the linearized calculations to the truncation order $M = 1$ in this paper.
	
	
	\section{Relaxation time approximation}
	\label{sec:RTA}
	In order to verify the results of the above calculations, it is useful to compare the results with a simple, analytic estimate. We apply the relaxation time approximation (RTA), in which the collision term is assumed to take a simple form
	\begin{align}
		\sum_{j=1}^{N_{\text{species}}}\mathcal{C}_{ij}^{(1)}[f_{i,\textbf{k}}] = -\frac{u_\mu k_i^\mu}{\tau} f^{(1)}_{i,\mathbf{k}} = -\frac{E_{i,\mathbf{k}}}{\tau} f^{(1)}_{i,\mathbf{k}}, \label{eq:RTAApprox}
	\end{align} 
	where $\tau$ is the relaxation time. The relaxation time can be interpreted as a global mean free-time between collisions of particles, and is an input parameter.
	
	\subsection{Diffusion coefficients in RTA}
	Applying the RTA to the linearized Boltzmann equation \eqref{eq:FinalLinBoltzEq} allows us to directly identify its analytical solution for the perturbation
	\begin{align}
	-\sum_{q\in\lbrace B,Q,S \rbrace}f^{(0)}_{i,\mathbf{k}} k_{i}^{\langle\mu\rangle} \nabla_\mu \alpha_{q} \left(\frac{E_{i,\mathbf{k}} n_{q}}{\epsilon + P_0} - q_i\right) &= -\frac{E_{i,\mathbf{k}}}{\tau} f^{(1)}_{i,\mathbf{k}} \nonumber \\
	& \Rightarrow \quad  f^{(1)}_{i,\textbf{k}} = \tau \sum_{q \in \lbrace B,Q,S \rbrace} \frac{k^{\langle \mu \rangle}_i}{E_{i,\textbf{k}}}   \left( \frac{E_{i,\textbf{k}} n_q}{\epsilon + P_0} - q_i \right) f^{(0)}_{i,\textbf{k}} \nabla_\mu \alpha_q .
	\end{align}
	The diffusion currents then take the form
	\begin{align}
		j^\mu_q &\equiv \sum_{i=1}^{N_{\text{species}}} q_i \int \mathrm{d}K_i k^{\langle \mu \rangle}_i f_{i,\textbf{k}}^{(1)} = \frac{\tau}{3} \sum_{q^\prime \in \lbrace B,Q,S \rbrace} \nabla^\mu \alpha_{q^\prime} \sum_{i=1}^{N_{\text{species}}} q_i \int \mathrm{d}K_i \frac{1}{E_{i,\textbf{k}}} \left(m_i^2 - E_{i,\textbf{k}}^2 \right) \left( \frac{E_{i,\textbf{k}} n_{q^\prime} }{\epsilon + P_0} - q^\prime_i \right) f^{(0)}_{i,\textbf{k}} \nonumber \\ 
		&\overset{!}{=} \sum_{q^\prime \in \lbrace B,Q,S \rbrace} \kappa_{qq^\prime} \nabla^\mu \alpha_{q^\prime},
	\end{align}
	and by direct comparison we arrive at the RTA expression for the diffusion coefficients:
	\begin{align}
		\kappa_{qq^\prime} = \frac{\tau}{3} \sum_{i=1}^{N_{\text{species}}} q_i \int \mathrm{d}K_i \frac{1}{E_{i,\textbf{k}}} \left(m_i^2 - E_{i,\textbf{k}}^2 \right) \left( \frac{E_{i,\textbf{k}} n_{q^\prime} }{\epsilon + P_0} - q^\prime_i \right) f^{(0)}_{i,\textbf{k}}. \label{eq:RTA}
	\end{align}
	This expression can also be written as
	\begin{align}
		\kappa_{qq^\prime} = \frac{\tau}{3} \left[ \sum_{i=1}^{N_{\text{species}}} q_i q^\prime_i \int \mathrm{d}K_i \frac{1}{E_{i,\textbf{k}}} \left(E_{i,\textbf{k}}^2 - m_i^2 \right)   f^{(0)}_{i,\textbf{k}}
										 +\sum_{i=1}^{N_{\text{species}}} \frac{n_{q^\prime}q_i }{\epsilon + P_0}\int \mathrm{d}K_i \left(m_i^2 - E_{i,\textbf{k}}^2 \right)    f^{(0)}_{i,\textbf{k}}\right], \label{eq:RTA_terms}	
	\end{align}						
	where the last integral gives the partial equilibrium pressure $P_{0i}$ of particle species $i$ that, in the Boltzmann gas, can be written as $P_{0i} = n_i T$, where $n_i$ is the total number density of the particle species. Thus, the expression for the diffusion coefficients becomes
	\begin{align}
		\kappa_{qq^\prime} = \frac{\tau}{3} \sum_{i=1}^{N_{\text{species}}} q_i q^\prime_i \int \mathrm{d}K_i \frac{1}{E_{i,\textbf{k}}} \left(E_{i,\textbf{k}}^2 - m_i^2 \right)   f^{(0)}_{i,\textbf{k}}
										 - \tau\frac{ T n_{q^\prime} n_{q}}{\epsilon + P_0}. \label{eq:RTA_final}	
	\end{align}
	Even if derived in the RTA, this expression allows us to identify the main features of the diffusion coefficients, in particular, its temperature dependence. We first note that in Eq. \eqref{eq:RTA_final} the symmetry of $\kappa_{qq^\prime}$ \cite{Onsager1931a,Onsager1931b} with respect to charge $q \leftrightarrow q^\prime$ is explicit. Moreover, we note that the charge combination $q_i q_i^\prime$ is the same for a particle and its corresponding anti-particle, so that the first term increases as the \emph{total} density of charge carriers increases, with the largest contribution coming from the lightest particle species that carries both charges $q$ and $q^\prime$. The last term is proportional to the net charge densities (particle minus antiparticle), and it becomes important when the net charge densities are comparable to the total charge density. Furthermore, the relaxation time is related to the inverse of the scattering rate, $\tau \sim 1/\Gamma_{\text{scatt.}} \sim 1/({n_{\mathrm{tot}} \sigma_{\mathrm{tot}}})$, and thus we deduce that the diffusion coefficients are suppressed by the scattering rate of the charged particles, which is strongly related to the total particle density of the medium with which they scatter. The dependence of $\kappa_{q q^\prime}$ on temperature and $\mu_B$ is discussed below in more detail when we show the results for hadron gas. 
	
	\subsection{Ultrarelativistic limit} 
	\label{sec:UltrarelLimit}
	In the ultrarelativistic limit, all coefficients for fixed chemical potentials, $\mu_q = \mathrm{const.}$, have the same asymptotic limit. In order to show this, we first make use of the fact that in the case of massless, classical particles, the thermodynamic integrals simplify
	\begin{align}
	\int \mathrm{d}K_i\, E_{i,\textbf{k}}^n f_{i,\textbf{k}}^{(0)} = g_i \frac{(n+1)!}{2\pi^2}T^{n+2}\exp(\alpha_i),
	\end{align}
	which leaves us with the expression for the diffusion coefficients in the massless case in the RTA
	\begin{align}
	\kappa_{qq^\prime} = -\frac{\tau}{3} \frac{T^3}{\pi^2} \sum_{i=1}^{N_{\text{species}}} g_i q_i \exp(\alpha_i) \left( \frac{3T n_{q^\prime}}{\epsilon + P_0} - q^\prime_i \right) . \label{eq:MasslessRTA}
	\end{align}
	For fixed chemical potentials, all mass and chemical scales can be neglected in the ultrarelativistic limit, since $m_i/T \rightarrow 0$ and $\alpha_i = \mu_i/T \rightarrow 0$ for all particle species. Because all thermal potentials vanish, all net charge densities $n_{q}$ also vanish in this limit. The high temperature limit follows  directly from the massless limit expression \eqref{eq:MasslessRTA} with $\exp(\alpha_i) \rightarrow 1$, and thus reads
	\begin{align}
	\kappa_{qq^\prime} = \frac{\tau}{3} \frac{T^3}{\pi^2} \sum_{i=1}^{N_{\text{species}}} g_i q_i q^\prime_i .
	\end{align}
	The relaxation time can be related to the scattering rate as
	\begin{align}
	\tau \sim \frac{1}{\Gamma_{\text{scatt.}}} \sim \frac{C}{n_{\text{tot}} \sigma_{\text{tot}}} , \label{eq:RelaxationTimeModel}
	\end{align}
	where $n_{\text{tot}} = \sum\limits_i \langle E_{i,\textbf{k}} \rangle_{i,0}$ is the total particle density, $\sigma_{\text{tot}}$ the total averaged cross section for the interaction between the constituents of the gas, and $C$ is a constant. For the massless case, we can then write that
	\begin{align}
	\tau T^3 = \frac{C \pi^2}{\sigma_{\text{tot}} \sum_i g_i \exp(\alpha_i)},
	\end{align}
	and in the limit introduced above, this simplifies to
	\begin{align}
	\tau T^3 = \frac{C \pi^2}{\sigma_{\text{tot}} \sum_i g_i}.
	\end{align}
	With this, the diffusion coefficients in the ultrarelativistic limit read:
	\begin{align}
	\kappa_{qq^\prime} = \frac{1}{3} \frac{C}{\sigma_{\text{tot}} \sum\limits_j g_j} \sum\limits_{i=1}^{N_{\text{species}}} g_i q_i q_i^\prime,
	\end{align}
	which is independent of any chemical potential. Furthermore, it becomes a constant if the total cross section, $\sigma_{\text{tot}}$, is constant, while, in the conformal limit (where the cross section must scale with the temperature as $\sigma_{\text{tot}} \sim 1/T^2$) $\kappa_{qq^\prime}/T^2$ becomes constant. These are properties that we also found to be true in the full linearized numerical evaluation.
	
	In recent publications \cite{Denicol2018a,Li2018}, the authors took the baryon diffusion coefficient of a massless QGP in RTA to be $\kappa_{B} = \frac{C_B}{T} n_{\mathrm{B}} \left( \frac{1}{3} \coth(\alpha_{\mathrm{B}}) - \frac{T n_{\mathrm{B}}}{\epsilon + P_0} \right)$, where the relaxation time was assumed to be $\tau = C_B/T$. This relation is also a special case of Eq.~\eqref{eq:MasslessRTA} for the case of a massless gas with a particle and a corresponding anti-particle species with baryon charge $B = \pm 1$:
	\begin{align}
	\kappa_{\mathrm{BB}} &\overset{\eqref{eq:MasslessRTA}}{=} -\frac{1}{3} \frac{C_B}{T} \frac{T^3}{\pi^2} g_S  \left[\exp(\alpha_{\mathrm{B}}) \left( \frac{3T n_{\mathrm{B}}}{\epsilon + P_0} - 1 \right) - \exp(-\alpha_{\mathrm{B}}) \left( \frac{3T n_{\mathrm{B}}}{\epsilon + P_0} + 1 \right)\right] \nonumber\\
	&= \frac{C_B}{T} \underbrace{2\frac{T^3}{\pi^2} g_S \sinh(\alpha_{\mathrm{B}})}_{\equiv \, n_{\mathrm{B}}} \left( \frac{1}{3} \coth(\alpha_{\mathrm{B}}) - \frac{T n_{\mathrm{B}}}{\epsilon + P_0} \right) = \frac{C_B}{T} n_{\mathrm{B}} \left( \frac{1}{3} \coth(\alpha_{\mathrm{B}}) - \frac{T n_{\mathrm{B}}}{\epsilon + P_0} \right).
	\end{align}
	
	\subsection{Validity of the relaxation time approximation}
	
\begin{figure}[h!]
		\centering
		\includegraphics[width=.7\textwidth]{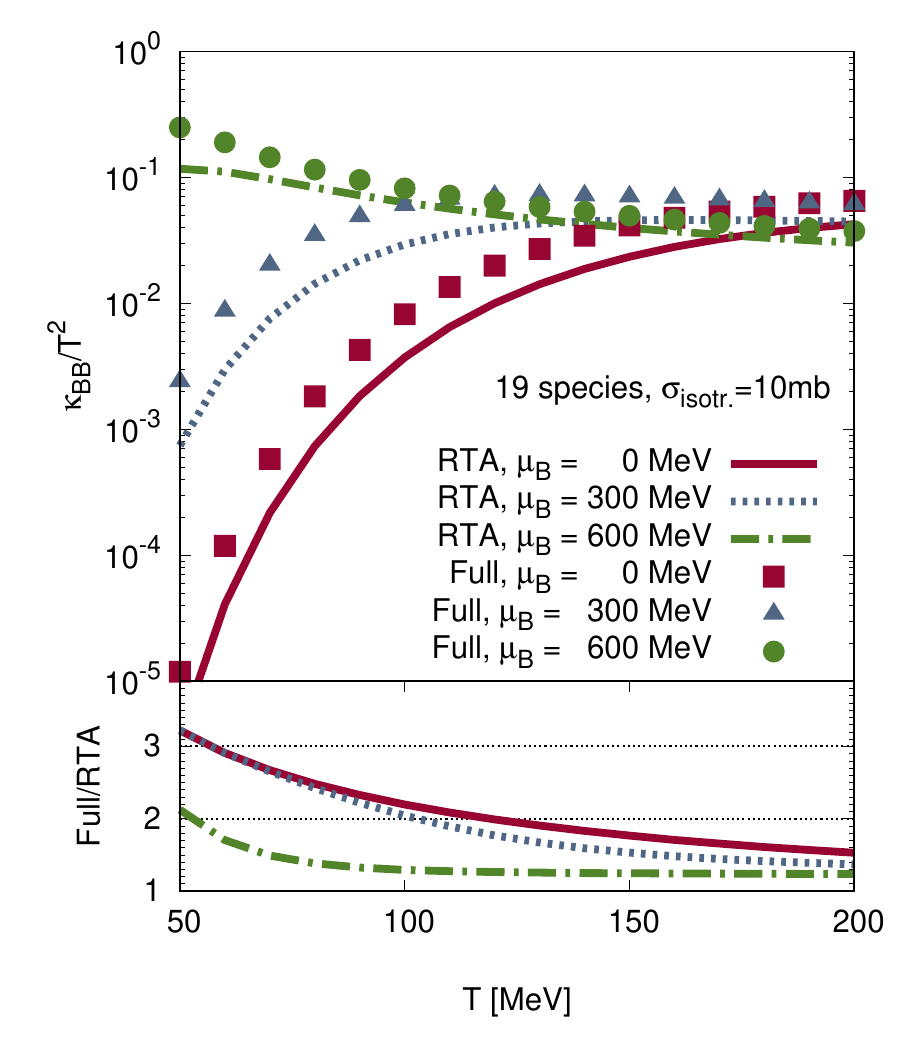}
		\caption{Top panel: We compare the complete linearized calculation of the baryon diffusion coefficient, $\kappa_{\mathrm{BB}}$, from Eq.~\eqref{eq:FinalLinearizedDiffu} (label 'Full' - full points) with assumed fixed isotropic cross section $\sigma_{\rm{tot}}$ to results from the relaxation time approximation (label 'RTA' - lines), Eq.~\eqref{eq:RTA}, where $ \tau_{\text{tr}}^{-1} = \frac{2}{3} n_{\mathrm{tot}} \sigma_{\rm{tot}} $. Results are presented for three values of baryon chemical potential ($\mu_{\mathrm{B}} = 0, \, 300, \, 600$ MeV) and vanishing electric and strangeness chemical potential, $\mu_{\mathrm{S}} = 0 = \mu_{\mathrm{Q}}$. Bottom panel: The ratio of RTA and full results.}
			\label{fig:KappaBOverT2_T_RTA_vs_FULL}
	\end{figure}

	In this section, we show that the relaxation time approximation retains the correct scaling behavior in temperature and baryon chemical potential for constant cross sections. In order to investigate when the RTA is applicable, we compute the baryon diffusion $\kappa_{\mathrm{BB}}$ for the lightest 19 hadron species (listed in Appendix \ref{sec:ParticleProperties}) with a constant isotropic cross section ($10~\mathrm{mb}$), using both the linearized collision term (later denoted as "full" in Fig. \ref{fig:KappaBOverT2_T_RTA_vs_FULL}), Eq. \eqref{eq:ExpandedLinColl}, and its relaxation time approximation (RTA), Eq. \eqref{eq:RTAApprox}. To this end, we employ the transport relaxation time $\tau_{\mathrm{tr}}$ (\eqref{eq:RelaxationTimeModel} with $C = \frac{3}{2}$),
	\begin{align}
		\tau^{-1}_{\mathrm{tr}} = n_{\mathrm{tot}} \sigma_{\mathrm{tr}} = \frac{2}{3} n_{\mathrm{tot}} \sigma_{\mathrm{tot}}. \label{eq:PlotDiff}
	\end{align}
	This form originates from the assumption of a constant differential cross section, $\frac{\mathrm{d}\sigma(\varphi,\vartheta)}{\mathrm{d}\varphi \mathrm{d}\vartheta} = \mathrm{const.}$, which is weighted at large scattering angles:
	\begin{align}
	\sigma_{\mathrm{tr}} \equiv \int\limits_0^{2\pi} \mathrm{d}\varphi \int\limits_0^{\pi} \mathrm{d}\vartheta \sin(\vartheta) \sin^2(\vartheta) \frac{\mathrm{d}\sigma(\varphi,\vartheta)}{\mathrm{d}\varphi \mathrm{d}\vartheta} = \frac{2}{3} 4\pi \frac{\mathrm{d}\sigma(\varphi,\vartheta)}{\mathrm{d}\varphi \mathrm{d}\vartheta} = \frac{2}{3} \sigma_{\mathrm{tot}},
	\end{align}
	where 
	\begin{align}
	\sigma_{\mathrm{tot}} \equiv \int\limits_0^{2\pi} \mathrm{d}\varphi \int\limits_0^{\pi} \mathrm{d}\vartheta \sin(\vartheta) \frac{\mathrm{d}\sigma(\varphi,\vartheta)}{\mathrm{d}\varphi \mathrm{d}\vartheta} = 4\pi \frac{\mathrm{d}\sigma(\varphi,\vartheta)}{\mathrm{d}\varphi \mathrm{d}\vartheta}.
	\end{align}
	The comparison is shown in Fig.~\ref{fig:KappaBOverT2_T_RTA_vs_FULL} for several values of baryon chemical potential $\mu_{\mathrm{B}}$. We note that the temperature dependence of the full calculation is reproduced well by the RTA evaluation. Additionally, we remark that the $\mu_{\mathrm{B}}$-dependence vanishes at high temperatures and that a ($1/T^2$)-scaling of $\kappa_{\mathrm{BB}}/T^2$ is achieved at very high temperatures (which is not shown in Fig.\ \ref{fig:KappaBOverT2_T_RTA_vs_FULL}), which we discussed in the last section for the case of constant cross sections. The full calculation deviates by a factor of $1-3$ from the RTA and improves with larger temperatures accordingly. Finally, we conclude that the RTA becomes more reliable at higher temperatures, but any quantitative study should retain the full collision term as we have done; especially if non-constant cross sections (which in general introduce additional dependencies on temperature and chemical potential) are present.

	
	\section{Diffusion coefficient matrix of a hadronic gas}
	\label{sec:HadronicGasResults}
	
\begin{figure}[h!]
		\centering
		\includegraphics[width=0.7\textwidth]{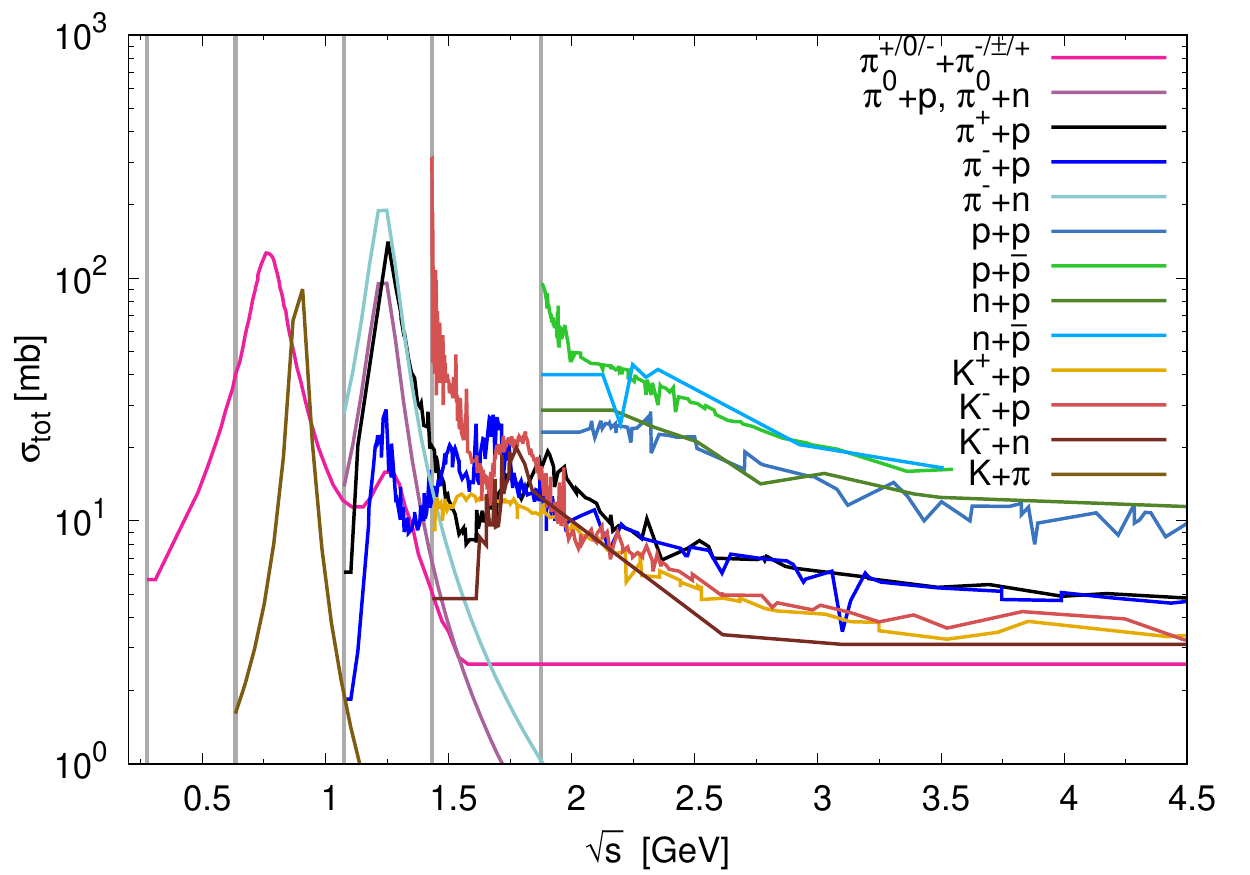}
		\caption{The isotropic resonance cross sections from the particle data book \cite{PDG}, GiBUU \cite{GiBUU} and SMASH \cite{Weil:2016zrk} that we use for the computation of the diffusion coefficients. All combinations of species not listed are approximated by constant values (see Appendix \ref{sec:ParticleProperties}). The grey bars denote the minimal $\sqrt{s}$ for the corresponding interaction. This plot was taken from Ref. \cite{Greif2017PRL}.}
		\label{fig:all_resonances_paper}
	\end{figure}

	We provide results for the diffusion coefficient matrix computed for a gas of hadrons and characterized by elastic binary hadron-hadron collision cross sections. We model the hadron gas using the most dominant mesons and baryons in a hot gas close to the QGP phase transition, that is, pions, kaons, nucleons, as well as lambda- and sigma-baryons (for particle properties see Appendix \ref{sec:ParticleProperties}). From the particle data group \cite{PDG}, we use all available elastic, isotropic cross sections and complement other theoretically-described resonant cross section from GiBUU \cite{GiBUU} and SMASH \cite{Weil:2016zrk}, as shown in Fig.~\ref{fig:all_resonances_paper}. All missing cross sections are approximated by the constant values taken from UrQMD \cite{UrQMD1,UrQMD2}, or (approximated from) GiBUU \cite{GiBUU}, as given in table ~\ref{tab:1} in Appendix \ref{sec:ParticleProperties}.

	In the following, we present and discuss results for calculations completed in this particular example of a hadronic system, where we assumed a temperature range of $T = 60$ MeV to $180$ MeV and a baryon chemical potential range of $\mu_{\mathrm{B}} = 0$ to $600$ MeV. The electric chemical potential is set to zero, $\mu_{\mathrm{Q}} = 0$, for simplicity. Furthermore, in this section, when we show the transport coefficients we always set the net strangeness density to be zero, $n_{\mathrm{S}}=0$ (as is expected to occur in the initial stages of heavy-ion collisions). This condition results in a strangeness chemical potential that cannot be zero, but must be a function of temperature and baryon chemical potential, $\mu_{\mathrm{S}} = \mu_{\mathrm{S}}(T,\mu_{\mathrm{B}})$. However, as we will see later, the cross-coupling between the diffusion currents can dynamically generate regions of non-zero net strangeness during the (fluid) dynamical evolution, even if it is initially zero. Therefore, in order to perform fluid dynamical simulations where the diffusion is taken fully into account, it is necessary to compute the full table of diffusion coefficients with arbitrary combinations of temperature and chemical potentials. Here, we show the positive baryon chemical potential range, however, we emphasize that there is in general no symmetry along the individual $\mu_q$ axes. The coefficients are only symmetric under point reflections: if the sign of \emph{all} chemical potentials are changed simultaneously. Further, we emphasize that there is no phase transition included in this model because this is a evaluation from (weakly-coupled) kinetic theory.
	
	Due to the systematic uncertainty in the cross sections, we vary all approximated constant values in Tab.~\ref{tab:1} in Appendix \ref{sec:ParticleProperties} by multiplying them by a factor $k=0.5,1,2$. We show this uncertainty of the diffusion coefficients by transparent bands in Figs.~\ref{fig:KappaBBOverT2_T_PDG} -- \ref{fig:KappaQQOverT2_T_PDG}. Furthermore, in Figs.~\ref{fig:KappaBBOverT2_T_Mu_PDG_3D} -- \ref{fig:KappaQQOverT2_T_Mu_PDG_3D} we show the full $T$ and 
	$\mu_{\mathrm{B}}$ dependence of the diffusion coefficients in 3D plots. 	
	
	As discussed in Sec.~\ref{sec:RTA}, the magnitude of the diffusion coefficients depends on both: the total and net density of the corresponding charge carriers, as well as their scattering rate. In turn, the scattering rate depends on the total particle density (with pion as the lightest hadron giving the largest contribution) and on the scattering cross section of the charge carriers. Many of the cross sections show quite a strong dependence on the particle collision energy (see Fig.~\ref{fig:all_resonances_paper}) and this also is somewhat reflected in the temperature dependence of the scattering rates. Thus, the temperature and chemical potential dependence of the diffusion coefficients is a result of the interplay between the energy and hadron type dependence of the scattering cross sections, as well as the temperature and chemical potential dependence of the hadron densities.

	\subsection{Baryon diffusion}
	
	First, we begin with the diffusion of baryon number, which concerns the diffusion coefficients $\kappa_{\mathrm{BB}}$, $\kappa_{\mathrm{BS}}$ and $\kappa_{\mathrm{BQ}}$ in the coefficient matrix. They measure the response of baryon number due to gradients in baryon-, strangeness- and electric-chemical thermal potential
	\begin{align}
		j^\mu_{\mathrm{B}} = \kappa_{\mathrm{BB}} \nabla^\mu \alpha_{\mathrm{B}} + \kappa_{\mathrm{BS}} \nabla^\mu \alpha_{\mathrm{S}} + \kappa_{\mathrm{BQ}} \nabla^\mu \alpha_{\mathrm{Q}} .
	\end{align}
	
\begin{figure}[h!]
		\begin{subfigure}{0.475\textwidth}
			\centering
			\includegraphics[width=\textwidth]{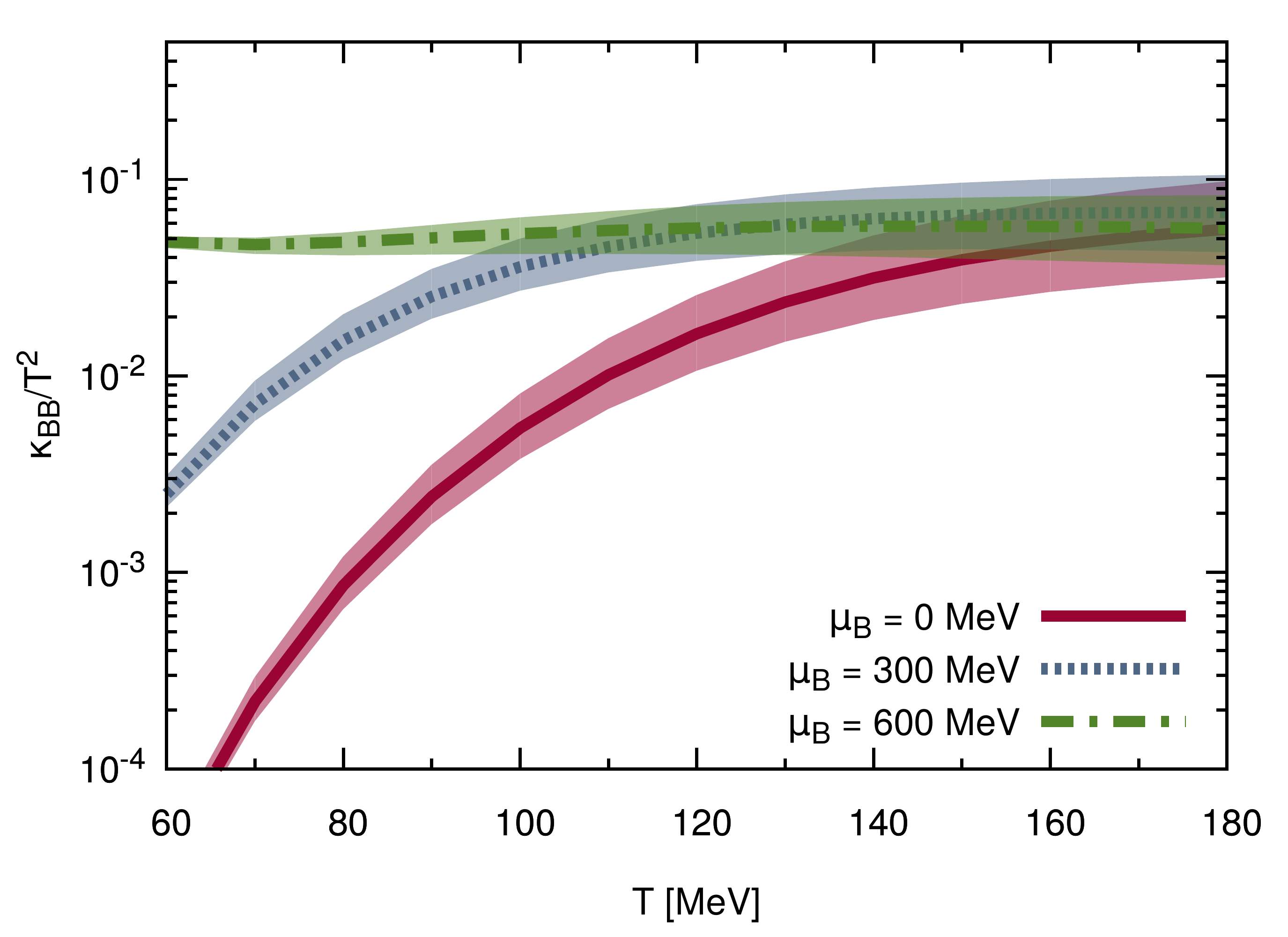}
			\caption{}
			\label{fig:KappaBBOverT2_T_PDG}
		\end{subfigure} \hfill
		\begin{subfigure}{0.475\textwidth}
			\centering
			\includegraphics[width=\textwidth]{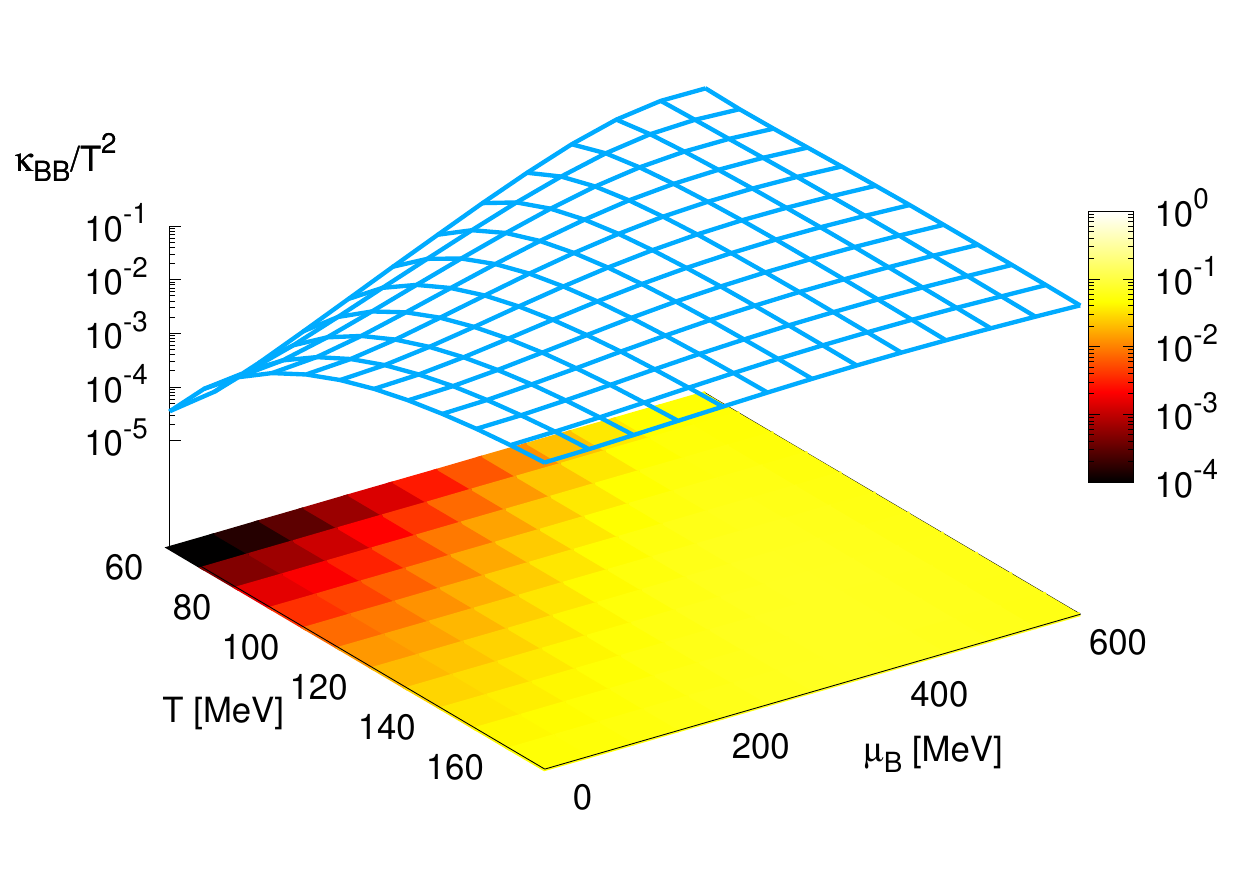}
			\caption{}
			\label{fig:KappaBBOverT2_T_Mu_PDG_3D}
		\end{subfigure}
		\caption{\textit{Left (a)}: Scaled baryon diffusion coefficient, $\kappa_{\mathrm{BB}}/T^2$, of a hadron gas with particle species and interactions listed in Appendix \ref{sec:ParticleProperties} and Figure \ref{fig:all_resonances_paper}, plotted in a temperature range $60$ to $180$ MeV and for baryon chemical potentials $\mu_{\mathrm{B}} = 0$, $300$ and $600$ MeV. We show bands, where the variation of the constant cross sections by a factor of $0.5$ and $2$. \textit{Right (b)}: 3D-plot of the same coefficient over temperature and baryon chemical potential. Both plots were evaluated in the case of vanishing net strangeness density, $n_{\mathrm{S}} = 0$, and $\mu_{\mathrm{Q}} = 0$, which implies that the strangeness chemical potential is a function of $\mu_{\mathrm{B}}$ and $T$.}
	\end{figure}

\begin{figure}[h!]
		\begin{subfigure}{0.475\textwidth}
			\centering
			\includegraphics[width=\textwidth]{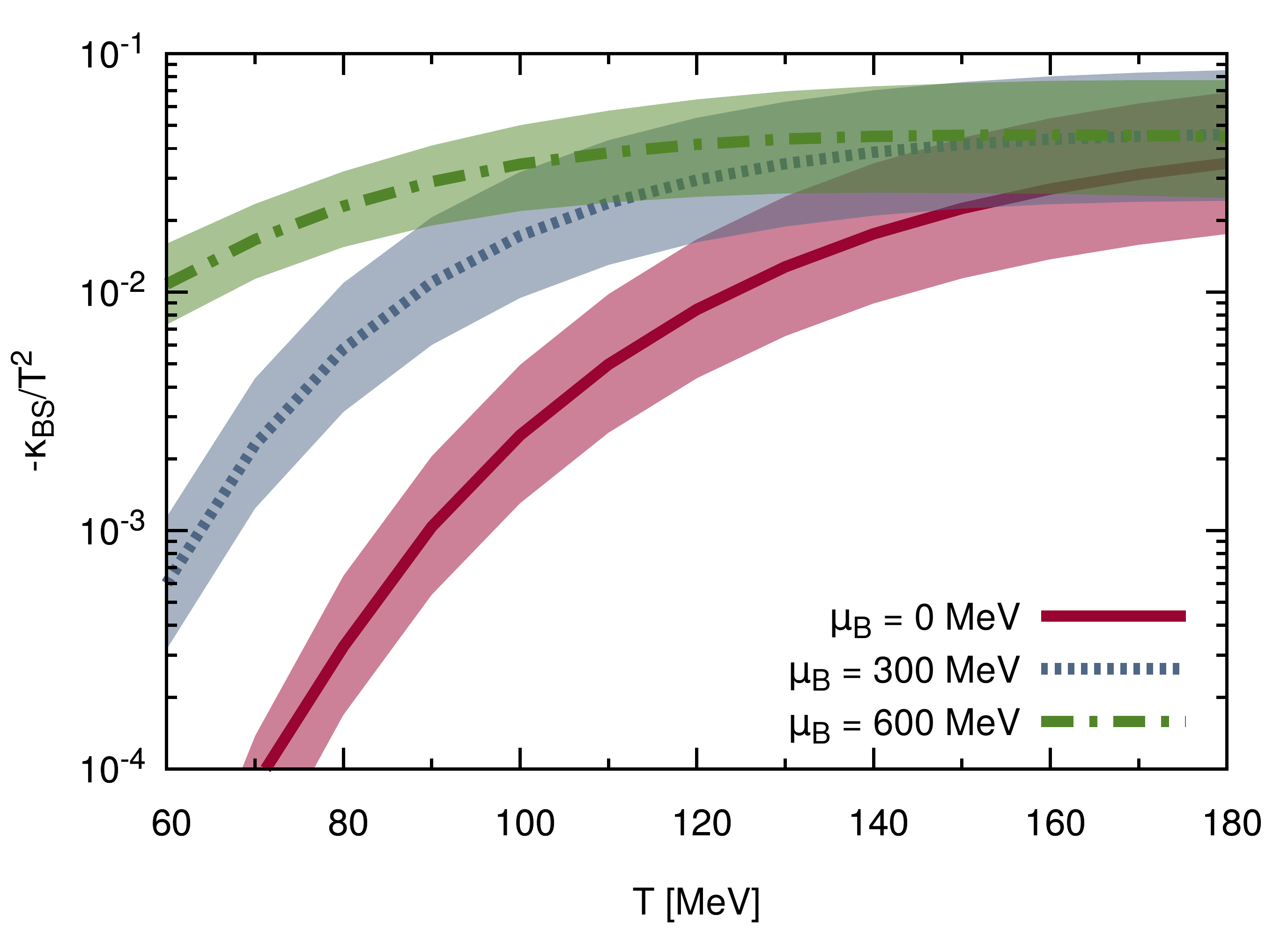}
			\caption{}
			\label{fig:KappaSBOverT2_T_PDG}
		\end{subfigure} \hfill
		\begin{subfigure}{0.475\textwidth}
			\centering
			\includegraphics[width=\textwidth]{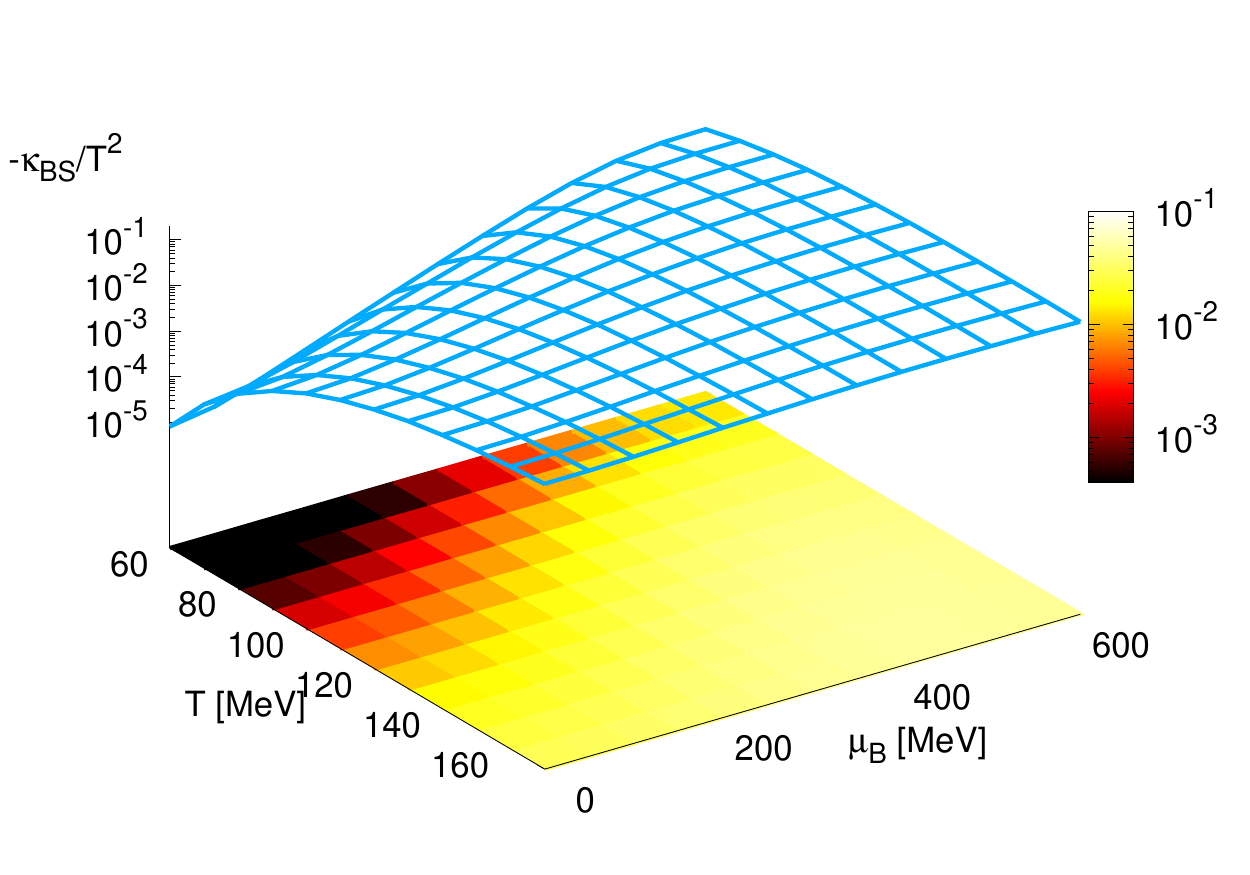}
			\caption{}
			\label{fig:KappaSBOverT2_T_Mu_PDG_3D}
		\end{subfigure}
			\caption{Same plots as in Figures \ref{fig:KappaBBOverT2_T_PDG} and \ref{fig:KappaBBOverT2_T_Mu_PDG_3D} for the scaled diffusion coefficient $\kappa_{\mathrm{SB}}/T^2$, which measures the diffusive coupling between strangeness and baryon number. Note that the coefficient is negative, and thus we plot $-\kappa_{\mathrm{SB}}$. \textit{Left (a)}: Band plot for over temperature and for different given $\mu_{\mathrm{B}}$. \textit{Right (b)}: 3D-plot of the same coefficient over temperature and baryon chemical potential.}
	\end{figure}

\begin{figure}[h!]
		\begin{subfigure}{0.475\textwidth}
			\centering
			\includegraphics[width=\textwidth]{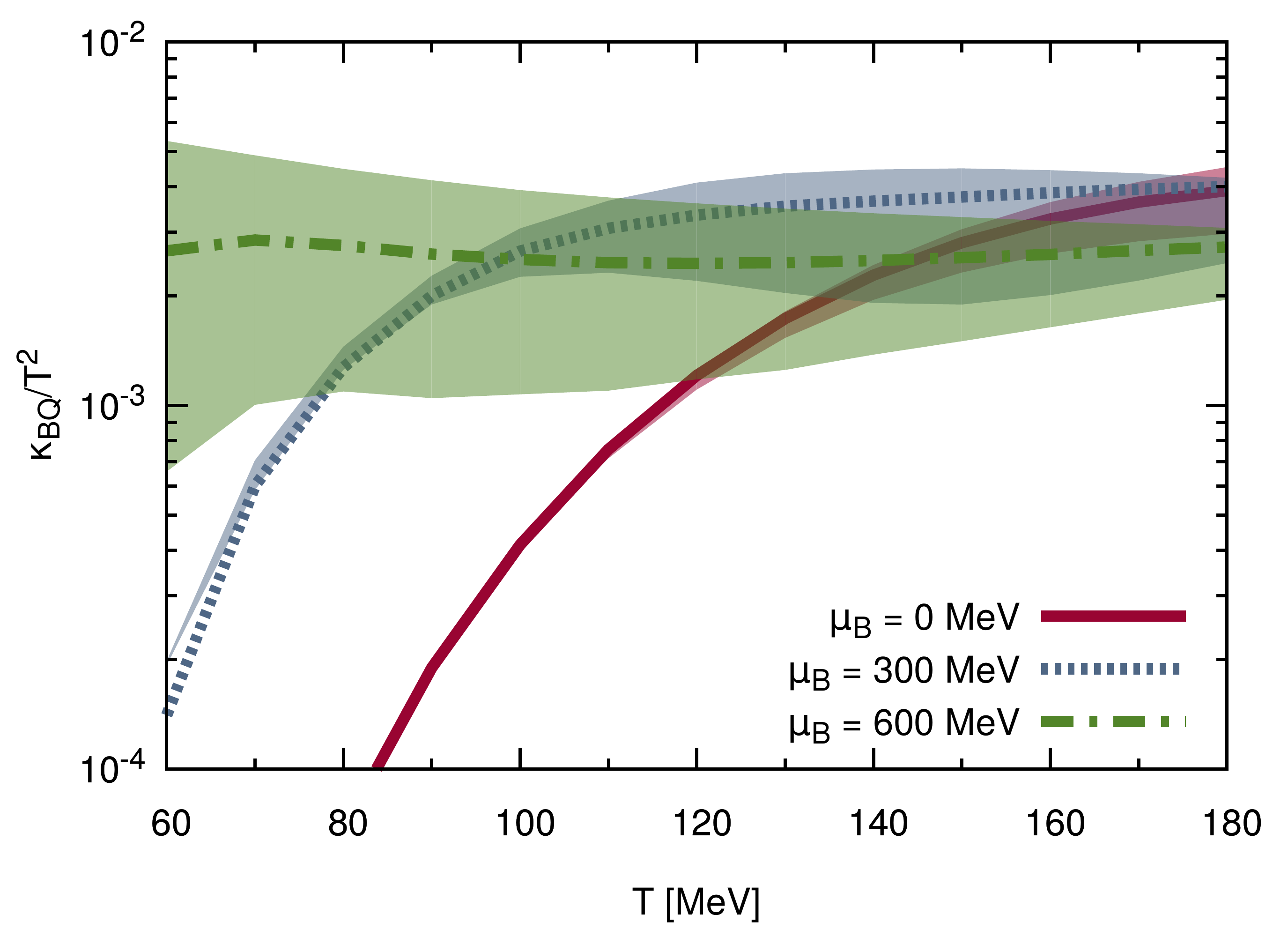}
			\caption{}
			\label{fig:KappaBQOverT2_T_PDG}
		\end{subfigure} \hfill
		\begin{subfigure}{0.475\textwidth}
			\centering
			\includegraphics[width=\textwidth]{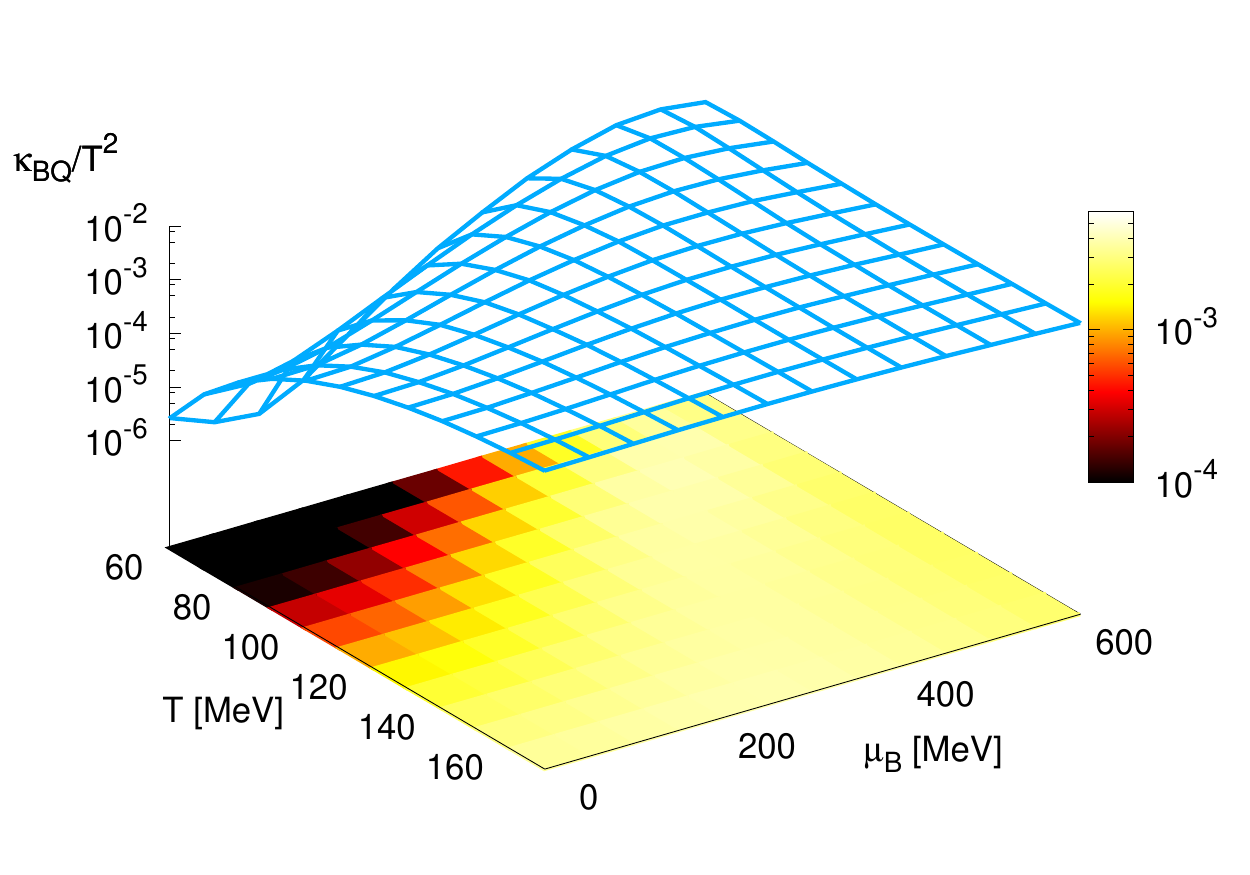}
			\caption{}
			\label{fig:KappaBQOverT2_T_Mu_PDG_3D}
		\end{subfigure}
			\caption{Same plots as in Figures \ref{fig:KappaBBOverT2_T_PDG} and \ref{fig:KappaBBOverT2_T_Mu_PDG_3D} for the scaled diffusion coefficient $\kappa_{\mathrm{BQ}}/T^2$, which measures the diffusive coupling between electric and baryon charges. \textit{Left (a)}: Coefficient plotted over temperature and for a variety of baryon chemical potentials with bands due to the variation of the constant cross sections. \textit{Right (b)}: 3D-plot of the same coefficient over temperature and baryon chemical potential.}
	\end{figure}

	All the diffusion coefficients in the baryon sector, which are shown in Figs.~\ref{fig:KappaBBOverT2_T_PDG}, \ref{fig:KappaBBOverT2_T_Mu_PDG_3D}, \ref{fig:KappaSBOverT2_T_PDG}, \ref{fig:KappaSBOverT2_T_Mu_PDG_3D}, \ref{fig:KappaBQOverT2_T_PDG}, and \ref{fig:KappaBQOverT2_T_Mu_PDG_3D}, display a rather strong dependence on temperature and baryon chemical potential. In particular, at $\mu_{\mathrm{B}}=0$, the magnitude of the coefficients increases by a factor of $\sim 10^4$, within the studied temperature range (note that the diffusion coefficients are divided by $T^2$ in the plots), and a similar increase is observed at lowest temperatures, within the studied $\mu_{\mathrm{B}}$ range. In both cases, the fast increase can be attributed to the rapid increase of the total baryon density. With nucleons being the lightest baryon number carriers, the baryon density is strongly suppressed by the Boltzmann mass factor $\exp(-m/T)$ and the relative increase of the baryon density with temperature is much faster than the increase of the total particle density, which is mainly determined by pions. Therefore, at $\mu_{\mathrm{B}}=0$, the increase of baryon density clearly wins over the increase in the scattering rate as temperature increases, and this results in the strong temperature dependence of the diffusion coefficients at low temperatures. At non-zero $\mu_{\mathrm{B}}$, the rapid increase of density with temperature is tamed by the fugacity factor $\exp(\mu_{\mathrm{B}}/T)$ and the temperature dependence becomes considerably milder at $\mu_{\mathrm{B}} = 600$ MeV.
	
	Similarly, the increase in $\mu_{\mathrm{B}}$ mainly affects the baryon density, with pion density being unaffected, and thus results in the fast increase of the diffusion coefficients with increasing $\mu_{\mathrm{B}}$. When $\mu_{\mathrm{B}}$ becomes sufficiently large, the baryon density is comparable to the total hadron density, and thus begins to affect the scattering rate, and the increasing scattering rate also limits the growth of the diffusion coefficients. At the same time, the net density of baryons also becomes comparable to the total density and this further limits the diffusion coefficient, see Eq.~(\ref{eq:RTA_final}).
	
	All the diffusion coefficients of the baryon sector ($\kappa_{\mathrm{BB}}$, $\kappa_{\mathrm{BS}}$ and $\kappa_{\mathrm{BQ}}$) display very similar behavior. This is due to the fact that the lightest hadron that contributes to these coefficients are nucleons for $\kappa_{\mathrm{BB}}$ and $\kappa_{\mathrm{BQ}}$, and hyperons for $\kappa_{\mathrm{BS}}$, with similar masses, and thus very similar behavior of densities. We further note that the qualitative behavior of $\kappa_{BB}$ is very similar in the test case with constant cross section (see Fig.~\ref{fig:KappaBOverT2_T_RTA_vs_FULL}). In this case, the energy and hadron type dependence of the cross sections only play a small role. The most visible difference is that with constant cross section, $\kappa_{\mathrm{BB}}/T^2$ at $\mu_{\mathrm{B}} = 600$ MeV actually decreases with increasing temperature, whereas this does not happen with more realistic cross sections.
	
	In the case of the baryon diffusion coefficient, $\kappa_{\mathrm{BB}}$, the shown bands demonstrate that the multiplicative factor in front of the constant cross sections changes the results more strongly at high temperatures, where the constant cross sections dominate the interactions in our study. Contrary to this, in the results for $\kappa_{\mathrm{BS}}$ we see that these bands have a large width over the whole temperature range. This is because most of the assumed interactions of hyperons were modeled with constant cross sections and these are the only charge carriers contributing to this particular coefficient due to the charge combination. We also note that $\kappa_{\mathrm{BS}}$ is negative due to the definition of the strangeness: hyperons with positive baryon number carry negative strangeness. This indicates a possible anti-correlation of baryon number and strangeness in dynamic simulations. In Fig.\ \ref{fig:KappaBQOverT2_T_PDG} we find for $\kappa_{\mathrm{BQ}}$ that throughout most of the temperature range, the given bands are narrow.

	\subsection{Strangeness diffusion}
	
\begin{figure}[h!]
		\begin{subfigure}{0.475\textwidth}
			\centering
			\includegraphics[width=\textwidth]{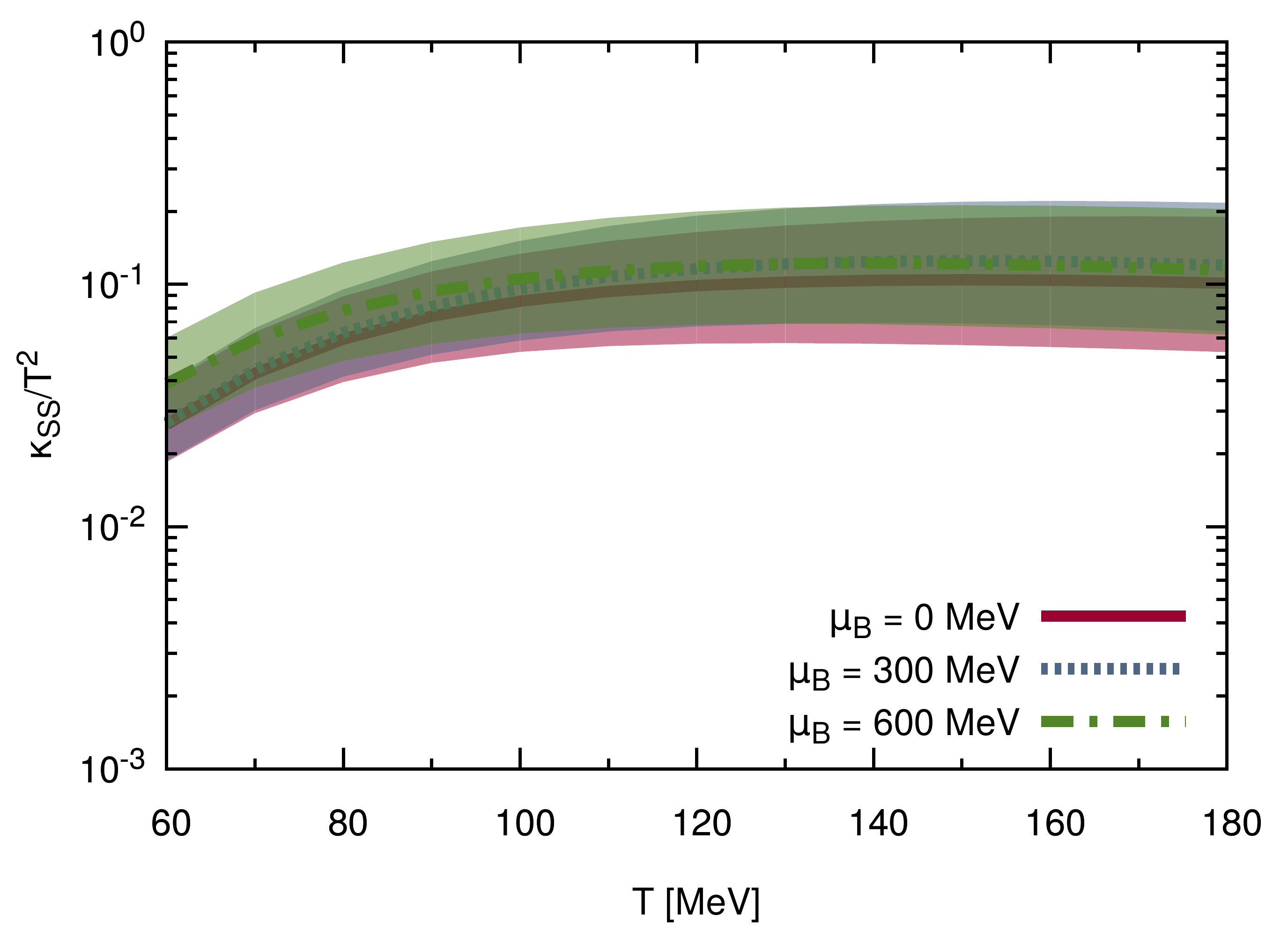}
			\caption{}
			\label{fig:KappaSSOverT2_T_PDG}
		\end{subfigure} \hfill
		\begin{subfigure}{0.475\textwidth}
			\centering
			\includegraphics[width=\textwidth]{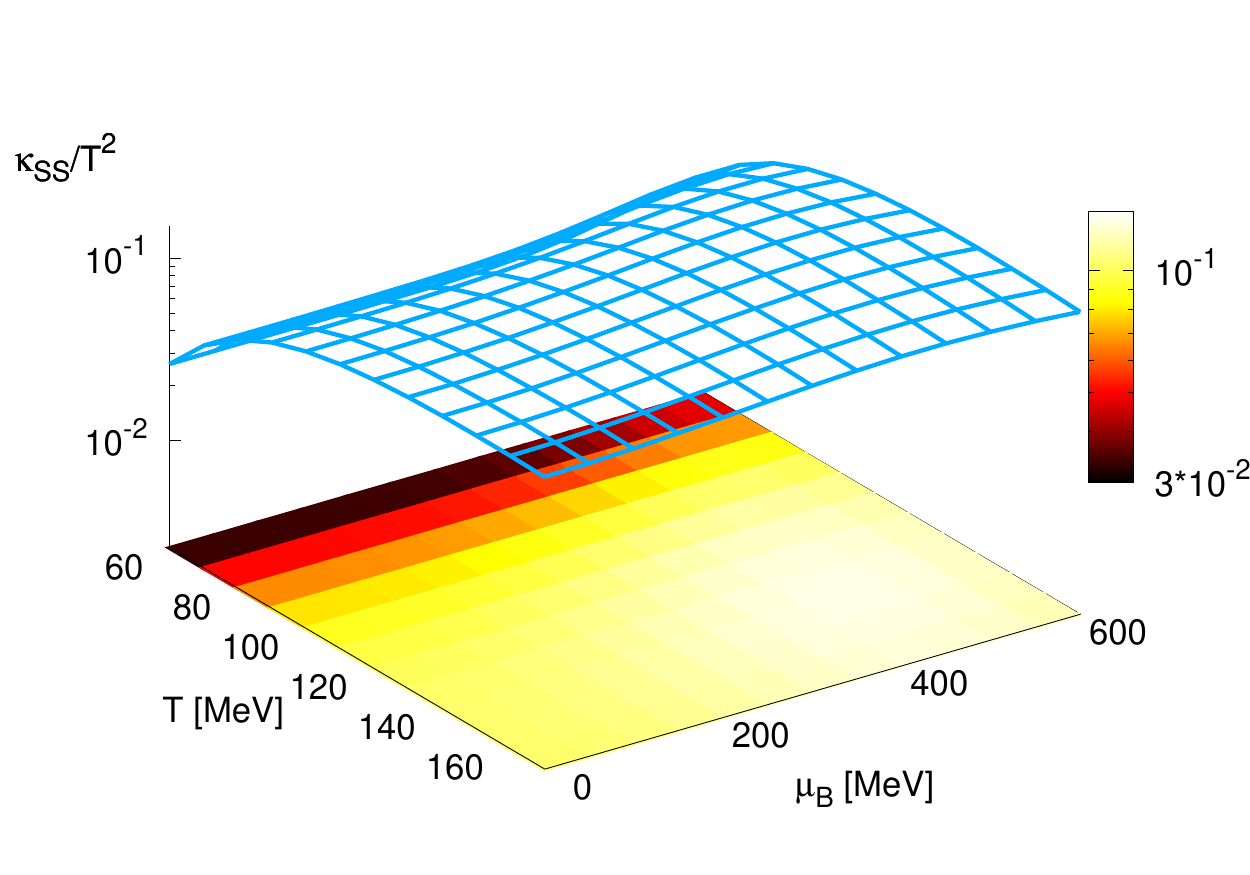}
			\caption{}
			\label{fig:KappaSSOverT2_T_Mu_PDG_3D}
		\end{subfigure}
		\caption{Same plots as in Figures \ref{fig:KappaBBOverT2_T_PDG} and \ref{fig:KappaBBOverT2_T_Mu_PDG_3D} for the scaled strangeness diffusion coefficient $\kappa_{\mathrm{SS}}/T^2$. \textit{Left (a)}: Similar to Fig. \ref{fig:KappaSBOverT2_T_PDG} the bands show a large width over the entire temperature range. \textit{Right (b)}: 3D-plot of the same coefficient over temperature and baryon chemical potential.}		
	\end{figure}

\begin{figure}[h!]
		\begin{subfigure}{0.475\textwidth}
			\centering
			\includegraphics[width=\textwidth]{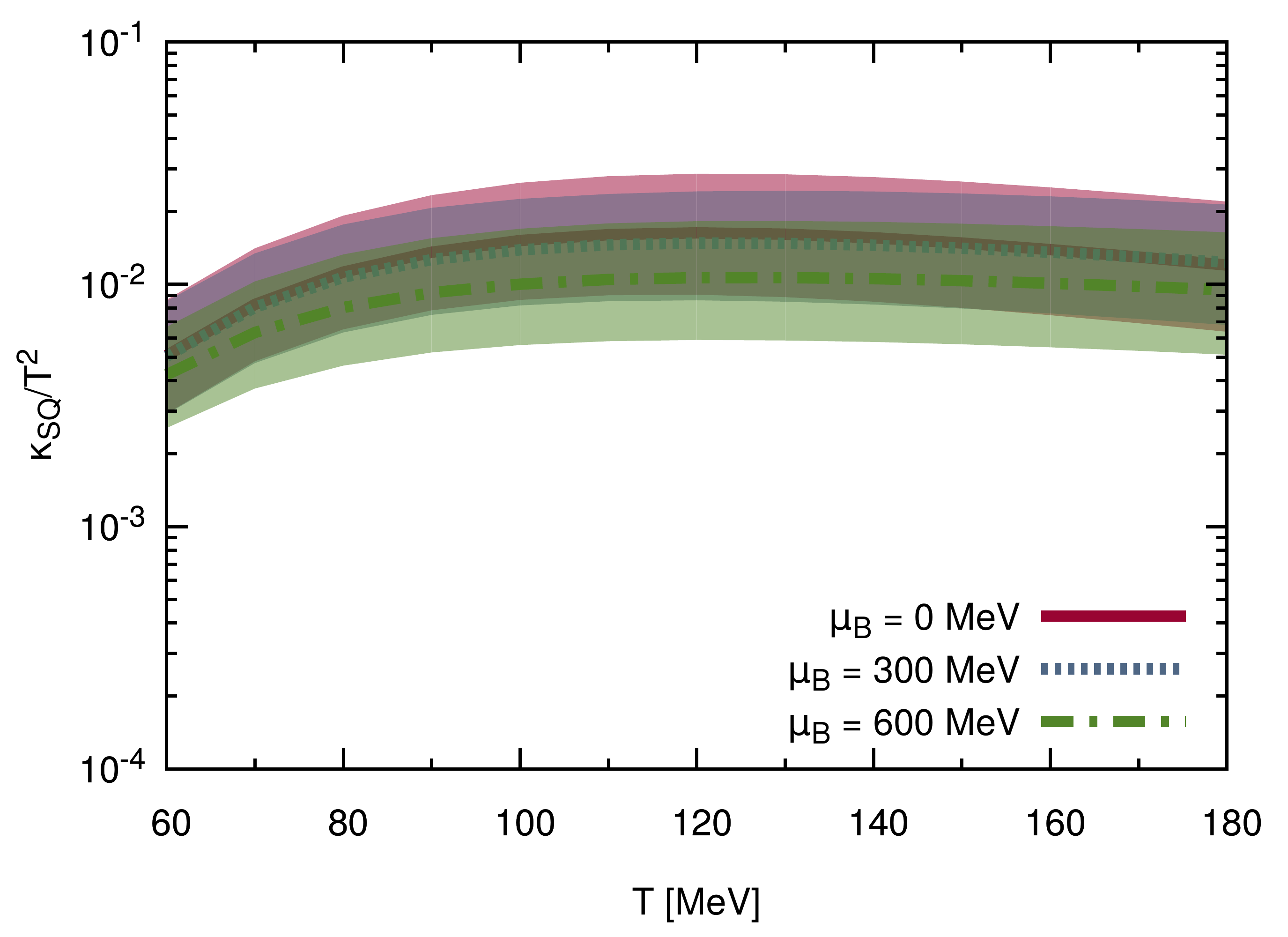}
			\caption{}
			\label{fig:KappaSQOverT2_T_PDG}
		\end{subfigure} \hfill
		\begin{subfigure}{0.475\textwidth}
			\centering
			\includegraphics[width=\textwidth]{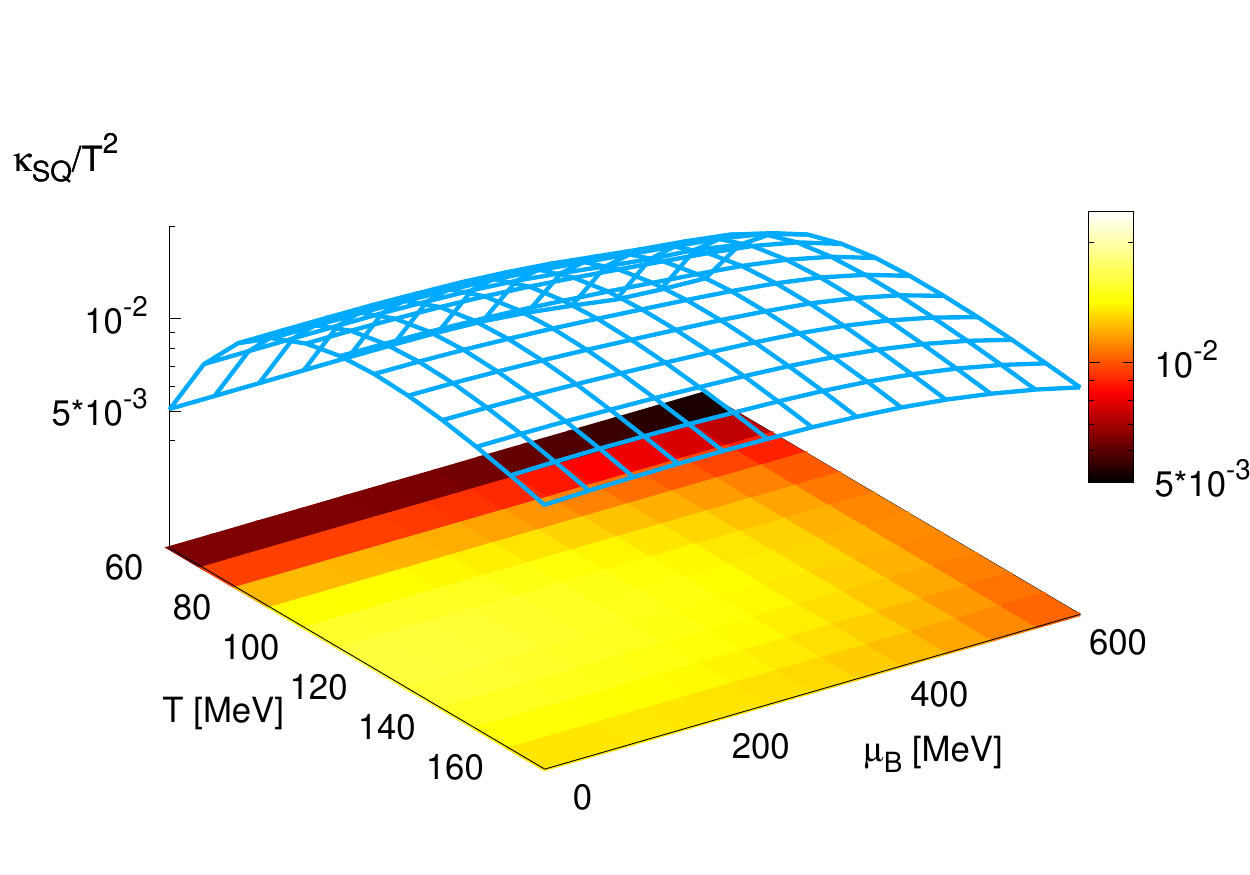}
			\caption{}
			\label{fig:KappaSQOverT2_T_Mu_PDG_3D}
		\end{subfigure}
		\caption{Same plots as in Figures \ref{fig:KappaBBOverT2_T_PDG} and \ref{fig:KappaBBOverT2_T_Mu_PDG_3D} for the scaled diffusion coefficient $\kappa_{\mathrm{SQ}}/T^2$, which measures the diffusive coupling between electric and strangeness charges. \textit{Left (a)}: Temperature plot of the coefficient shown for different values of $\mu_{\mathrm{B}}$ and the already introduced bands in the constant cross sections. \textit{Right (b)}: 3D-plot of the same coefficient over temperature and baryon chemical potential.}
	\end{figure}

	The diffusion of strangeness is characterized by the coefficients $\kappa_{\mathrm{SS}}$, $\kappa_{\mathrm{SB}}$ and $\kappa_{\mathrm{SQ}}$ via
	\begin{align}
	j^\mu_{\mathrm{S}} = \kappa_{\mathrm{SS}} \nabla^\mu \alpha_{\mathrm{S}} + \kappa_{\mathrm{SB}} \nabla^\mu \alpha_{\mathrm{B}} + \kappa_{\mathrm{SQ}} \nabla^\mu \alpha_{\mathrm{Q}} .
	\end{align}
	The $\kappa_{\mathrm{SB}}$-coefficient was already discussed because the diffusion matrix is symmetric and $\kappa_{\mathrm{BS}} = \kappa_{\mathrm{SB}}$ \cite{Onsager1931a,Onsager1931b}. For both of the remaining coefficients, 
	$\kappa_{\mathrm{SS}}$ and $\kappa_{\mathrm{SQ}}$, shown in Figs.~\ref{fig:KappaSSOverT2_T_PDG}, \ref{fig:KappaSSOverT2_T_Mu_PDG_3D}, \ref{fig:KappaSQOverT2_T_PDG}, and \ref{fig:KappaSQOverT2_T_Mu_PDG_3D}, the lightest hadron that carries strangeness, or both strangeness and electric charge, is the kaon. Similarly to the baryons, the increase in temperature leads to an increase in total strangeness density compared to the total density determining the scattering rate, but in this case the effect on the diffusion coefficient is much weaker since kaons are significantly lighter than the lightest baryons. Further, there is no significant dependence 
	on the baryon chemical potential since the kaons do not carry any baryon charge.
	
	\newpage
	
	\subsection{Electric diffusion}
	
\begin{figure}[h]
		\begin{subfigure}{0.475\textwidth}
			\centering
			\includegraphics[width=\textwidth]{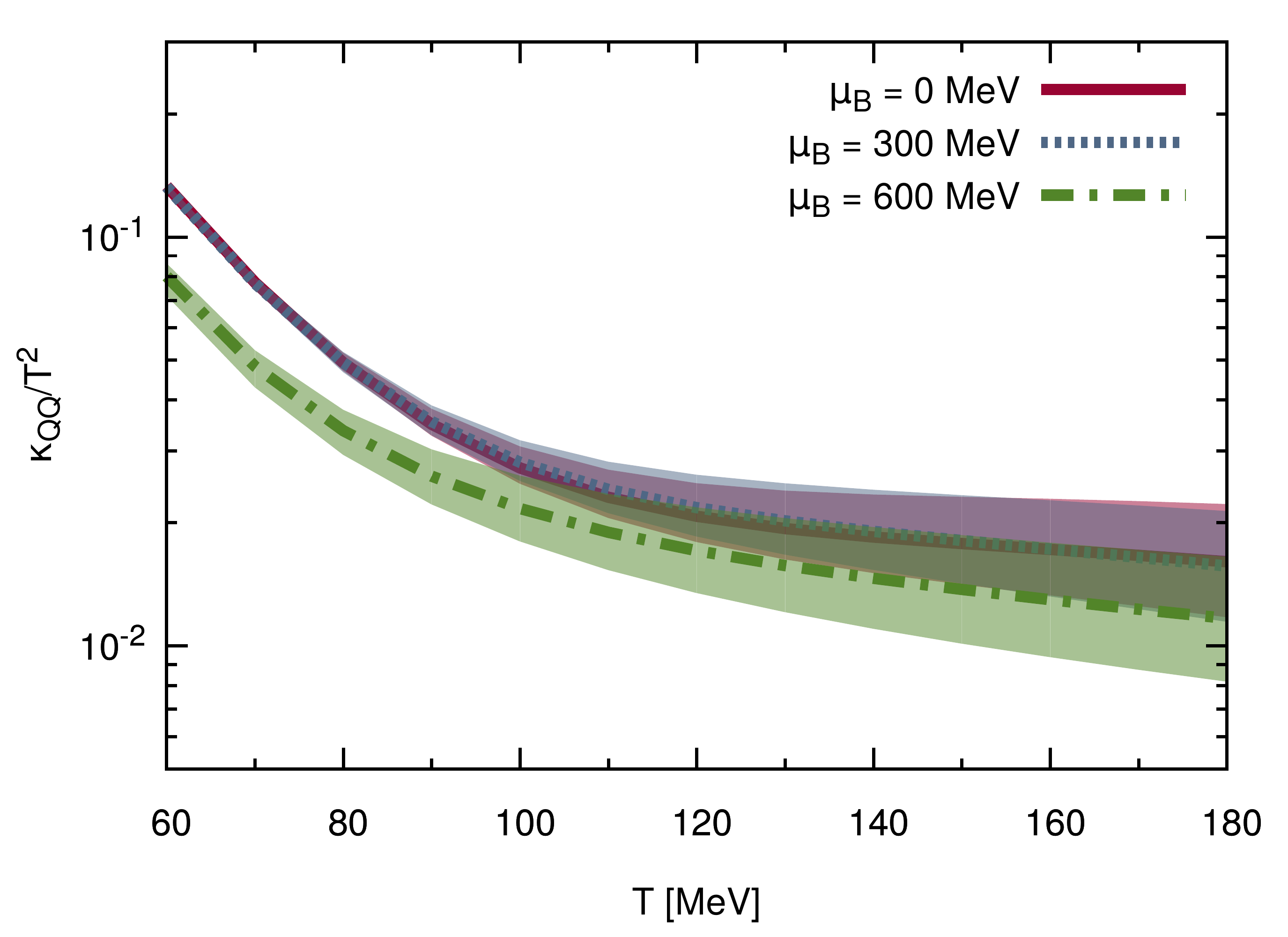}
			\caption{}
			\label{fig:KappaQQOverT2_T_PDG}
		\end{subfigure} \hfill
		\begin{subfigure}{0.475\textwidth}
			\centering
			\includegraphics[width=\textwidth]{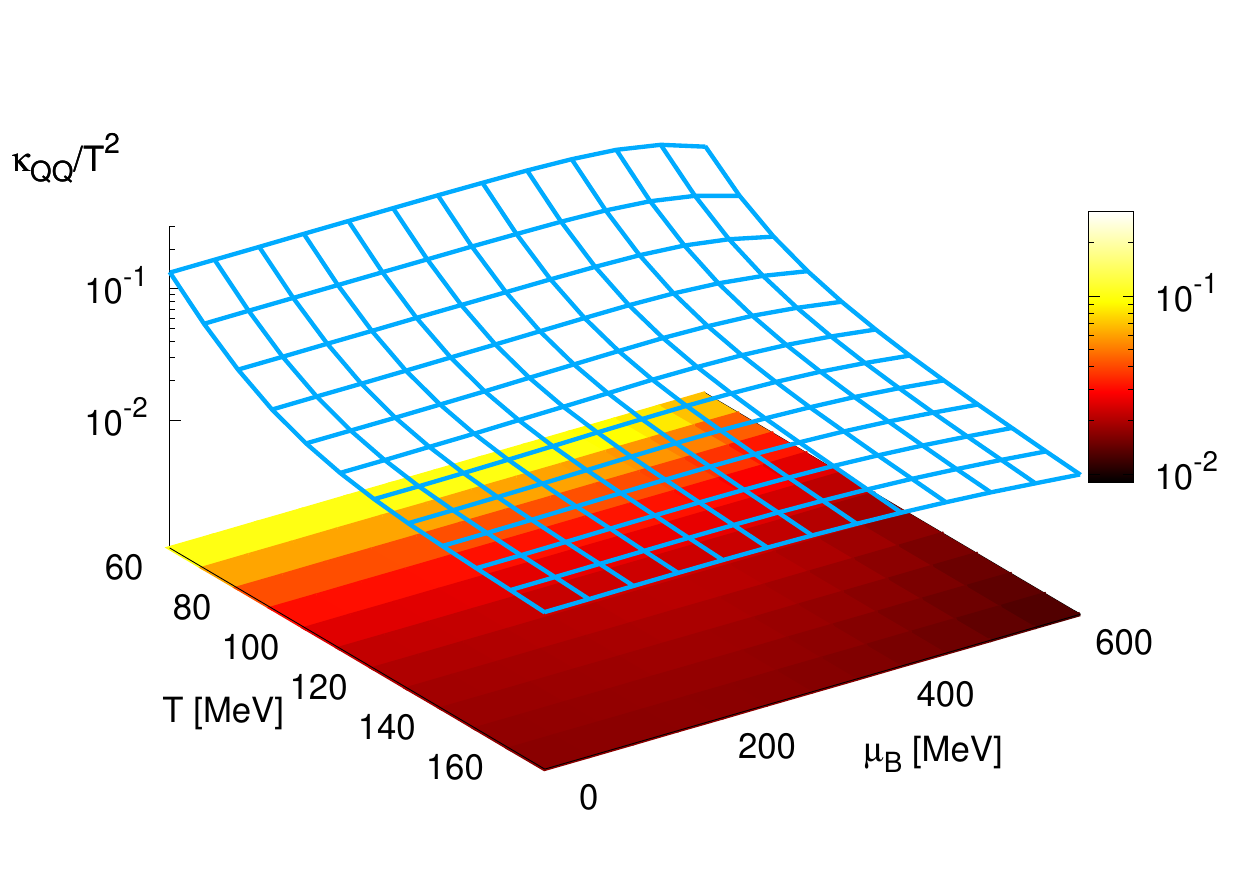}
			\caption{}
			\label{fig:KappaQQOverT2_T_Mu_PDG_3D}
		\end{subfigure}
		\caption{Same plots as in Figs.\ \ref{fig:KappaBBOverT2_T_PDG} and \ref{fig:KappaBBOverT2_T_Mu_PDG_3D} for the scaled electric diffusion coefficient $\kappa_{\mathrm{QQ}}/T^2$. \textit{Left (a)}: Coefficient plotted over temperature and for a variety of baryon chemical potentials with bands due to the variation of the constant cross sections.  \textit{Right (b)}: 3D-plot of the same coefficient over temperature and baryon chemical potential.}
	\end{figure}

	The response of electric charge due to gradients in all thermal charge potentials, $\alpha_q$, is measured by the coefficients $\kappa_{\mathrm{QQ}}$, $\kappa_{\mathrm{QB}}$ and $\kappa_{\mathrm{QS}}$
	\begin{align}
		j^\mu_{\mathrm{Q}} = \kappa_{\mathrm{QQ}} \nabla^\mu \alpha_{\mathrm{Q}} + \kappa_{\mathrm{QB}} \nabla^\mu \alpha_{\mathrm{B}} + \kappa_{\mathrm{QS}} \nabla^\mu \alpha_{\mathrm{S}} .
	\end{align} 
	The only coefficient left to discuss is $\kappa_{\mathrm{QQ}}$, which is shown in the same manner as before in Figs.\ \ref{fig:KappaQQOverT2_T_PDG} and \ref{fig:KappaQQOverT2_T_Mu_PDG_3D}. Contrary to $\kappa_{\mathrm{QB}}$, $\kappa_{\mathrm{QQ}}$ again shows no significant $\mu_{\mathrm{B}}$-dependence, since the most dominant charge carriers, the pions, do not carry any baryon charge. The fact that the lightest electric-charge carriers are also the lightest hadrons results in the fact that the total density of charge carriers grows at the same rate as the total hadron density, which in turn, determines the scattering rate. In this case, $\kappa_{\mathrm{QQ}}$ depends very weakly on temperature, and the shown temperature dependency in Fig.~\ref{fig:KappaQQOverT2_T_PDG} is from the $1/T^2$ scaling.
	
	We remark that for $\mu_{\mathrm{B}} = 0$, the electric diffusion coefficient, $\kappa_{\mathrm{QQ}}$, coincides with the electric conductivity calculated in Ref. \cite{Greif:2016skc}, $\kappa_{\mathrm{QQ}}(\mu_{\mathrm{B}} = 0) = T \sigma_{\mathrm{el}}(\mu_{\mathrm{B}} = 0)$. The similarity of electric conductivity and diffusion (or in the Eckart frame \cite{Eckart1940} the heat conductivity) is the manifestation of the Wiedemann-Franz law \cite{WiedemannFranz1853}. Similar to the electric conductivity, the electric diffusion coefficient also decreases strongly with temperature and only shows a mediocre dependence on $\mu_{\mathrm{B}}$ \cite{Greif:2016skc}. This is because the dominant electric charge carriers are the pions, but a significant amount of baryonic species also contribute to the electric diffusion current. Similar to Fig. \ref{fig:KappaBBOverT2_T_PDG}, the band widths vary strongly with increasing temperatures where the constant cross sections dominate the interactions.
	
	\newpage
	
	\section{The diffusion matrix for quark gluon plasma}
	\label{sec:massless}
	
	The computation of the hadronic diffusion coefficients presented above can only be extended in temperature up to $T \sim 160$ MeV where the transition to QGP is expected to happen. In order to extend the computation to higher temperatures, we need to complement the hadronic part by calculating the diffusion matrix also for the QGP. Here, we consider a toy model for the QGP where it is described as a massless gas of quarks and gluons undergoing isotropic elastic binary collisions. To this end, we take gluons and the three lightest quark flavors u, d, and s, together with their antiparticles. The degeneracy factors accounting for the spin and color degrees of freedom are $g=6$ for quarks and $g=16$ for gluons. 
	
	The magnitude of the diffusion coefficients is then determined by the collision cross section, which we must specify, and we do so in a very simplistic manner in order to get a first estimate. We can either fix the total cross section to a constant value, e.g.\ $\sigma_{\rm tot} = 10$ mb, or set the shear viscosity to entropy ratio $\eta/s$ to a fixed value. The latter choice is more suitable, since shear viscosity of the QGP is often extracted from experiment making this assumption. The shear viscosity is given by $\eta=2\epsilon/(5n_0 \sigma_{\mathrm{tot}})$ and the entropy density in chemical equilibrium is $s=4n_0$ \cite{Xu:2007ns,Bouras:2009nn}. Using the theoretical minimum $\eta/s=1/(4\pi)$ \cite{Kovtun:2004de}, the total isotropic cross section can be fixed to $\sigma_{\rm tot}\approx0.716/T^2$ \cite{Xu:2007ns,Bouras:2009nn}.  
	
	The quarks carry baryon number, strangeness and electric charge\footnote{Up-quark: $B = +1/3$, $S = 0$, $Q = +2/3$, Down: $B = +1/3$, $S = 0$, $Q = -1/3$, Strange: $B = +1/3$, $S = -1$, $Q = -1/3$ and corresponding anti-quarks. Gluon: $B = 0$, $S = 0$, $Q = 0$}, and the gluons contribute to the diffusion coefficients, mainly through the scattering rate. In the case of $s$-independent, isotropic cross sections, the diffusion coefficients scale with the total cross section $\sigma_{\mathrm{tot}}$. As an example, we give the massless limit of the complete diffusion coefficient matrix at vanishing chemical potential, $\mu_{q} = 0$ for $q \in \lbrace \mathrm{B}, \mathrm{Q}, \mathrm{S} \rbrace$:
	\begin{align}
	\begin{pmatrix}
	\kappa_{\mathrm{BB}} & \kappa_{\mathrm{BQ}} & \kappa_{\mathrm{BS}} \\  \kappa_{\mathrm{QB}} & \kappa_{\mathrm{QQ}} & \kappa_{\mathrm{QS}} \\ \kappa_{\mathrm{SB}} & \kappa_{\mathrm{SQ}} & \kappa_{\mathrm{SS}}
	\end{pmatrix}(\mu_q = 0) \approx 
	\frac{1}{\sigma_{\rm{tot}}} \begin{pmatrix}
	0.0345 & 0.0 & -0.0345 \\  0.0 & 0.0063 & 0.0105 \\ -0.0345 & 0.0105 & 0.1036
	\end{pmatrix}.
	\end{align}
	As discussed in Section \ref{sec:UltrarelLimit}, in the case of constant cross sections and $\mu_i=0$, the diffusion coefficients $\kappa_{qq^\prime}$ are also constant in the ultrarelativistic limit and the scaled coefficients $\kappa_{qq^\prime}/T^2$ therefore scale with inverse temperature squared. Contrary to this, in the conformal limit where $\eta/s = \text{const.}$ and $\sigma_{\rm tot}\sim T^{-2}$ the scaled coefficients are constant over temperature. At non-zero $\mu_{q}$, all the coefficient acquire temperature dependence through the Boltzmann factors $\exp(\mu_q/T)$ in the densities. However, at fixed $\mu_{q}$ the temperature dependence vanishes at large temperatures.
	
\begin{figure}
		\centering
		\includegraphics[width=0.95\textwidth]{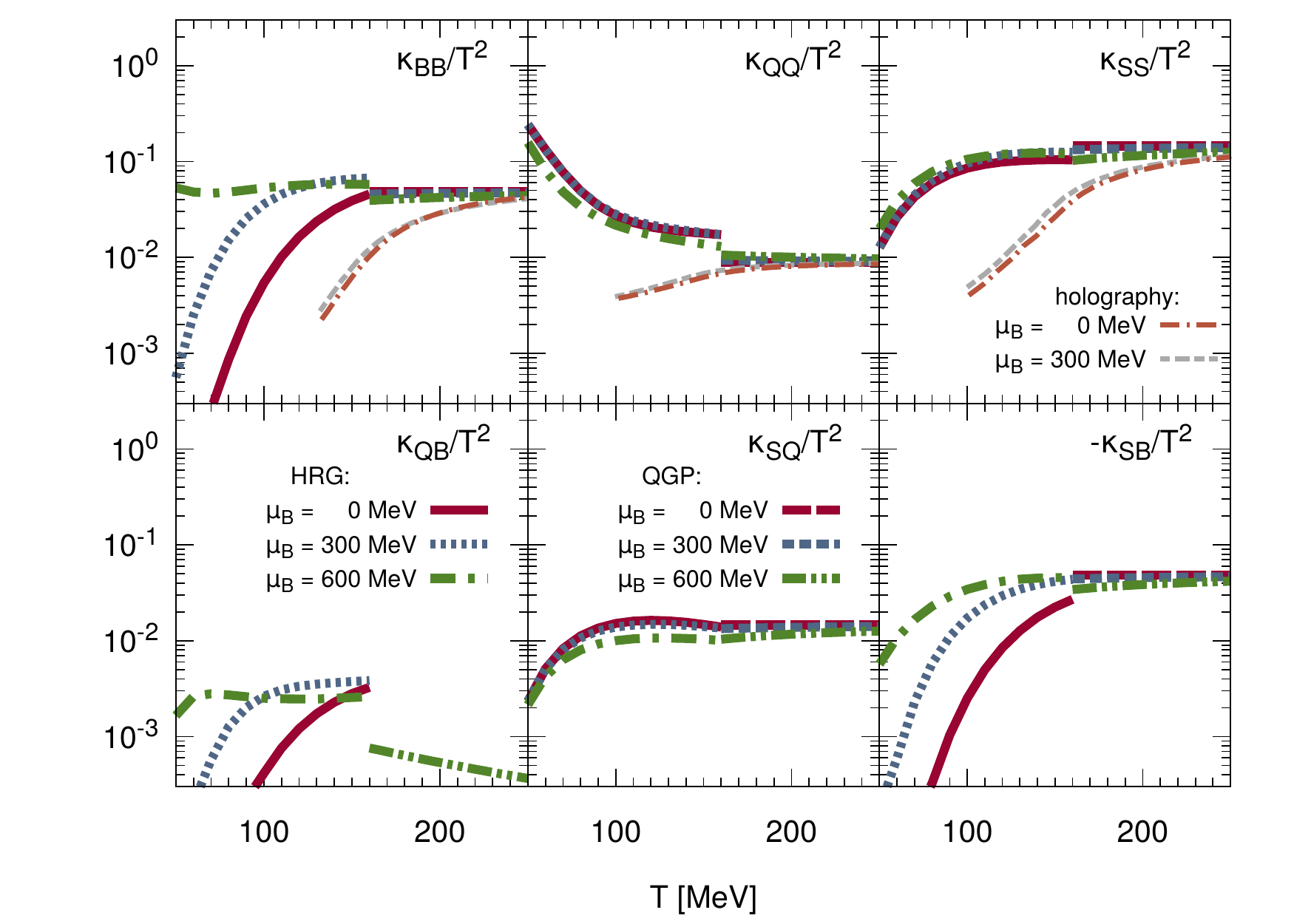}
		\caption{Complete diffusion coefficient matrix plotted over temperature and for the baryon chemical potentials $\mu_{\mathrm{B}} = 0$, $300$ and $600$ MeV. We show results for the assumed hadronic system in the temperature range $T=50$ to $160$ MeV and for the simple QGP model for fixed $\eta/s = 1/4\pi$ for temperatures above $160$ MeV. We compare to the holographic results achieved in Ref.~\cite{Rougemont:2015ona}. This plot was taken from Ref.~\cite{Greif2017PRL}.}
		\label{fig:final_plot}
	\end{figure}

	In Fig.~\ref{fig:final_plot} we plot the full diffusion coefficient matrix, where we show all acquired results for the hadronic gas already discussed in Section \ref{sec:HadronicGasResults} and also the results for the massless simple QGP model. We show results by fixing $\mu_{\mathrm{B}} = 0$, $300$ and $600$ MeV, the electric chemical potential to zero, $\mu_{\mathrm{Q}} = 0$, and also the net strangeness density to zero $n_{\mathrm{S}} = 0$ as in the hadronic case. We then simply compare and present our results for the two models in one summarizing plot and switch the model at 160 MeV temperature\footnote{We again emphasize that there is no phase transition included in this approach.}. We already note that on the QGP side there is very little dependence on temperature and baryon chemical potential, especially at large temperatures, as expected.
	
	Surprisingly, for $\mu_{\mathrm{B}} = 0$ the coefficients for the two different models almost match at $T=160$ MeV, where the phase transition or the smooth crossover would normally occur. The only exception seems to be the $\kappa_{\mathrm{BQ}}$, where there is a large discrepancy between the hadronic model and the QGP model. In the latter case, the coefficient vanishes at $\mu_{\mathrm{B}} = 0$ since the generated currents of the quarks \text{exactly} cancel out the net flow of electric charge due to symmetry. However, we speculate that the hadronic results for $\kappa_{\mathrm{BQ}}$ will decrease in magnitude if more hadronic particles are included in the computation, and will thus reduce the discrepancy between both models. 
	
	We further compare to the holographic results for the diagonal entries of the diffusion coefficient matrix from Ref.~\cite{Rougemont:2015ona} (grey dashed and dash-dot-dotted lines). These results approach the conformal limit at high temperatures and only show moderate $\mu_{\mathrm{B}}$ dependence, but the overall shape and magnitude seems to be qualitatively consistent with our results. However, the results for the simple QGP model with fixed $\eta/s = 1/4\pi$ coincides with the conformal limit at very high temperatures, as it should. We note that the $\kappa_{\mathrm{SB}}$-coefficient has the same magnitude as the baryon diffusion coefficient, $\kappa_{\mathrm{BB}}$, but is negative. We further remark that the strangeness diffusion coefficient is the largest coefficient in magnitude. 
	
	We found that the off-diagonal entries of the diffusion coefficient matrix can reach similar magnitudes to the diagonal coefficients (which are usually considered). We therefore would expect significant corrections to the diffusion currents due to the mixing of charge types compared to approaches when parts of the diffusion coefficient matrix are neglected. Nevertheless, the phenomenological consequences are still not known. In the following, we take a first step in this direction and investigate the influence of the full diffusion matrix in a one-dimensional fluid dynamics approach.

	\section{Hydrodynamic Evolution}
	\label{sec:Hydro}
	In the last sections we evaluated and discussed the diffusion coefficient matrix for a simple hadronic and (massless) partonic system and showed that in the chosen basis of charge definitions there are non-vanishing off-diagonal contributions arising from the fact that hadrons and partons can carry several different charges. The goal of this chapter is to provide initial investigations of its implications with the help of relativistic fluid dynamics. After providing a short review of our framework, we present the first results for the dynamic evolution of a system with multiple conserved charges. Here, we assume the same hadronic system as presented in Chapter \ref{sec:HadronicGasResults} as an example. More sophisticated studies will follow in the future.
	
	\subsection{Transient dissipative relativistic fluid dynamics}
	
	The foundation of fluid dynamics is the exact conservation of energy, momentum and the net quantum numbers (or charges) $q$. In the same fashion as in the last sections, we assume conserved baryon number $B$, strangeness $S$ and electric charge $Q$. The local conservation equations of energy, momentum and net charge $q$ can then be expressed in general (curved) spacetime as 
	\begin{align}
		0 = T^{\mu\nu}_{~;\mu} \equiv \partial_\mu T^{\mu\nu} + \Gamma^{\mu}_{~\mu \alpha} T^{\alpha \nu} + \Gamma^{\nu}_{~\mu \alpha} T^{\mu \alpha}, \quad 0 = N^\mu_{q,~;\mu} \equiv \partial_\mu N^\mu_{q} + \Gamma^\mu_{~\mu\alpha} N^\alpha_q ,
	\end{align} 
	where we introduced the covariant derivative, $(\cdots)_{;\mu}$, and the Christoffel symbols of the second kind, $\Gamma^{\mu}_{\alpha\beta}$.
	The central assumption of fluid dynamics is that the evolution of the fluid is taking place close to local equilibrium. This holds as long the characteristic microscopic scales of the system - e.g. the mean free-path of the particles - are sufficiently small compared to the (dominating) characteristic macroscopic scales of the system. Secondly, the dissipative corrections of the fluid dynamic tensors must be small in comparison to the equilibrium quantities. Both requirements are quantified by introducing the Knudsen numbers, $\mathrm{Kn}$, which are defined as the ratios of the microscopic and macroscopic scales, and the inverse Reynolds numbers $\mathrm{Rn}^{-1}$, which are defined as the ratios of the magnitude of the dissipative quantities (e.g. the diffusion currents), as well as the corresponding primary hydrodynamic fields (e.g. the local net charge densities). It is often argued that if both measures are small,
	\begin{align}
		\mathrm{Kn} \ll 1, \quad \mathrm{Rn}^{-1} \ll 1,
	\end{align}
	fluid dynamics is applicable. However, it was recently shown that in some situations the applicability of fluid dynamics extends even up to $\mathrm{Kn} \sim 1$ \cite{Gallmeister2018}. 
	
	In order to close the set of fluid dynamic equations, one needs to introduce additional equations of motion for the dissipative quantities. Following the approach of transient fluid dynamics in DNMR (Denicol-Niemi-Moln\'{a}r-Rischke) theory \cite{Denicol2012}, the equations of motion for the bulk viscous pressure, the diffusion currents, and the shear-stress tensor are introduced. The source terms responsible for the generation of dissipation in these equations are expanded in orders of Knudsen numbers and inverse Reynolds numbers under the assumption that they are sufficiently small so that the higher order contributions can be neglected. 
	
	For these first investigations where we want to examine the impact of the off-diagonal terms in the diffusion coefficient matrix, we only expand the source term to first order in the Knudsen numbers. The transient equations of motion then read \cite{Denicol2012}:
	\begin{align}
		\tau_\Pi\, \mathcal{D}\Pi + \Pi &= -\zeta \theta + \mathcal{O}(\mathrm{Kn}^2, \mathrm{Rn}^{-2}, \mathrm{Kn}\mathrm{Rn}^{-1}), \\
		\tau_q\, \Delta^{\mu}_{~\nu}\mathcal{D}j^{\nu}_{q} + j^\mu_{q} &= \sum_{q^\prime} \kappa_{qq^\prime} \nabla^\mu \alpha_{q^\prime} + \mathcal{O}(\mathrm{Kn}^2, \mathrm{Rn}^{-2}, \mathrm{Kn}\mathrm{Rn}^{-1}), \\
		\tau_\pi\, \Delta^{\mu\nu}_{\alpha\beta}\mathcal{D}\pi^{\alpha\beta} + \pi^{\mu\nu} &= 2\eta \sigma^{\mu\nu} + \mathcal{O}(\mathrm{Kn}^2, \mathrm{Rn}^{-2}, \mathrm{Kn}\mathrm{Rn}^{-1}),
	\end{align}
	where we introduced the bulk viscosity $\zeta$, the shear viscosity $\eta$ and accounted for the diffusion coefficient matrix $(\kappa_{qq^\prime})$. Further, $\mathcal{D} A^{\mu_1 \dots \mu_{\ell}} \equiv u^\alpha A^{\mu_1 \dots \mu_{\ell}}_{~;\alpha}$ is the comoving time derivative. The first order source terms correspond with the source terms from Navier-Stokes-Fourier theory \cite{landau1959course,Eckart1940}. We note that the Navier-Stokes terms do not contain any direct cross-couplings between the dissipative fields. In the following, we neglect bulk and shear, $\Pi = \pi^{\mu\nu} = 0$, and focus on the diffusion without viscous corrections. Thus, the only dissipative equations of motion we consider in this work are the equations for the net diffusion currents,
	\begin{align}
		\tau_q\, \Delta^\mu_{~\nu} \mathcal{D} j^{\nu}_{q} + j^\mu_{q} &= \sum_{q^\prime} \kappa_{qq^\prime} \nabla^\mu \alpha_{q^\prime} .
	\end{align}
	In order to solve these fluid dynamic equations of motion, we rewrite the set of equations in an appropriate manner ( see Appendix \ref{sec:ExplicitFluidDynamics}) and use the numerical solver SHASTA \cite{BorisBook,Molnar2009a}. For the sake of simplicity, we only assume longitudinal dynamics in a hyperbolic (1+1)D-geometry characterized by the proper time, $\tau \equiv \sqrt{t^2 - z^2}$, and the spacetime rapidity, $\eta_s \equiv \mathrm{arctanh}\left( z/t \right)$. We then solve Equations \eqref{eq:Ttautau}, \eqref{eq:Ttaueta}, \eqref{eq:Ntau} and \eqref{eq:DiffEta} numerically, and use Eqs. \eqref{eq:energdens} and \eqref{eq:chargedens} to infer the LRF quantities. Furthermore, we solve Eq. \eqref{eq:velocityeta} with Newtons secant algorithm in order to find the velocity. Please note that all equations in Appendix \ref{sec:ExplicitFluidDynamics} are already given without any viscous corrections.
	
	\subsection{Equation of state}
	
	In order to close the set of fluid dynamics equations, we need to impose an equation of state, $P_0\left(T, \mu_{\mathrm{B}}, \mu_{\mathrm{Q}}, \mu_{\mathrm{S}} \right)$. In the non-interacting hadron gas it is straightforward to compute thermodynamic quantities as a function of $T$ and $\mu_q$, 
	\begin{align}
	\epsilon_{\mathrm{eq}} \equiv \epsilon_{\mathrm{eq}}(T, \mu_{\mathrm{B}}, \mu_{\mathrm{Q}}, \mu_{\mathrm{S}}), \quad n_{q, \mathrm{eq}} \equiv n_{q, \mathrm{eq}}(T, \mu_{\mathrm{B}}, \mu_{\mathrm{Q}}, \mu_{\mathrm{S}}).
	\end{align}
	However, in fluid dynamics the natural variables are energy- and net charge densities, and we need to invert these relations numerically in order to obtain the pressure, temperature and the chemical potentials as a function of $\epsilon$ and $n_{q}$.
	Here we assume the same classical hadronic system as presented in Chapter \ref{sec:HadronicGasResults}, and thus the single-particle distribution function is of Maxwell-Juettner type \eqref{eq:MaxwellJuettner}, and the thermodynamic quantities can be expressed as,
	\begin{align}
		\epsilon_{\mathrm{eq}} \equiv \sum_{i=1}^{N_{\text{species}}} \Big\langle E^2_{i,\textbf{k}} \Big\rangle_{i,0}, \quad n_{q,\mathrm{eq}} \equiv \sum_{i=1}^{N_{\text{species}}} q_{i} \Big\langle E_{i,\textbf{k}} \Big\rangle_{i,0} ~ \text{for} ~ q \in \lbrace \mathrm{B}, \mathrm{Q}, \mathrm{S} \rbrace, \quad P_{0,\mathrm{eq}} \equiv \frac{1}{3}\sum_{i=1}^{N_{\text{species}}} \Big\langle E^2_{i,\textbf{k}} - m_i^2 \Big\rangle_{i,0}.  
	\end{align}
	We note that the equation of state constructed in this way is consistent with the equilibrium state in the computation of the diffusion matrix.

	\subsection{Results}
	
	In order to obtain some understanding of the diffusive interplay between the multiple conserved charges and the importance of the diffusion coefficient matrix, we simulate the dynamics of the hadronic system presented in Chapter \ref{sec:HadronicGasResults}. \newline
	
	\paragraph{Case study}\quad \\
	
	For the sake of simplicity, we consider only two conserved charges in the system, the net baryon number, and net strangeness by setting the electric chemical potential to zero, $\mu_{\mathrm{Q}} = 0$. We set simple initial conditions and consider four different configurations of the diffusion coefficient matrix of the system:
	\begin{itemize}
		\item \textit{Case 1}: No diffusion; all the diffusion coefficients are set to zero, $\kappa_{qq^\prime} = 0$.
		\item \textit{Case 2}: Baryon diffusion only; the only non-vanishing coefficient is $\kappa_{\mathrm{BB}}$, which is taken from the above-mentioned evaluation. This case is usually assumed in other works \cite{Denicol2018a,Li2018}. $\kappa_{\mathrm{BB}}$ is computed with the linear response method as described in the first part of this paper in the relevant range of temperature and chemical potentials. The only restriction implied is that the electric chemical potential vanishes, $\mu_{\mathrm{Q}} = 0$. 
		\item \textit{Case 3}: Off-diagonal entries neglected; all the off-diagonal entries of the coefficient matrix are artificially set to zero. Note that the only off-diagonal coefficient is $\kappa_{\mathrm{SB}} = 0$. All the diagonal coefficients are again taken from the above-mentioned calculation.
		\item \textit{Case 4}: Full diffusion matrix; the complete diffusion coefficient matrix of the system is considered.
	\end{itemize}
	We assume simple transversally homogeneous initial conditions for a heavy ion collision at small collisional energies, which is entirely in the hadronic phase and suffers large longitudinal gradients in net baryon number. In all of the above-introduced cases, the system is initialized at proper time $\tau_0 = 2$ fm/c, with a homogeneous temperature of 160 MeV and a double-gaussian profile in initial net baryon number density,
	\begin{align}
		n_{\mathrm{B},\mathrm{ini}} = n_{B,\mathrm{max}} \cdot \left[ \exp\left(-\frac{\left( \eta_s - \eta_{s,0} \right)^2}{R_0^2}\right) + \exp\left(-\frac{\left( \eta_s + \eta_{s,0} \right)^2}{R_0^2}\right) \right],
	\end{align}
	where $\eta_{s,0} = 1.0$, $n_{B,\mathrm{max}} = 0.5 \, \mathrm{fm^{-3}}$ and $R_0 = 0.5$. Furthermore, we set the initial net strangeness density to zero everywhere, $n_{\mathrm{S}} = 0$, and as usual, the initial fluid velocity is zero, $u^\mu = 0$ \footnote{Note that $u^\mu = 0$ in hyperbolic coordinates corresponds to $v_\perp = \sqrt{(v^x)^2 + (v^y)^2} = 0$ and $v^z = z/t$ in Cartesian coordinates, and therefore the perpetual longitudinal expansion is accounted for. Moreover, $n_{\mathrm{S}} = 0$ accounts for the fact that in the collisions of nucleons there is no initial net strangeness in the collision region.}. From these specifications, the energy density is calculated from the equation of state, which results in a non-homogeneous profile. This implies that besides the diffusion, the dynamics of the system are also determined by gradients in pressure. Therefore, even in the non-diffusive case (Case 1), the baryon number is transported with the flow of the system. \newline 
	
	\paragraph{Description} \quad \\
	
\begin{figure}[h]
			\centering
			\includegraphics[width=0.95\textwidth]{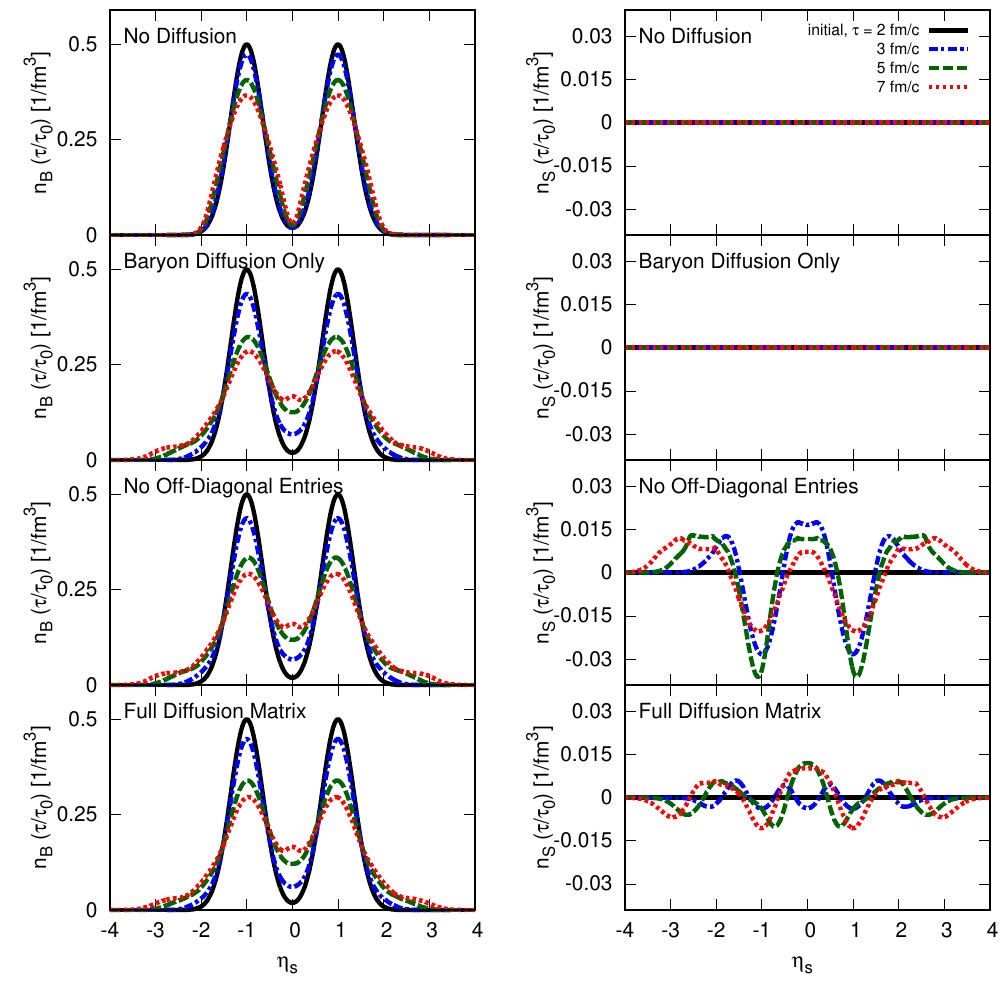}
			\caption{Longitudinal fluid dynamic evolution of the net baryon number (left plots) and net strangeness (right plots) multiplied by the Bjorken factor $\tau/\tau_0$ of a classical, hadronic system, with 19 assumed particle species (see Appendix \ref{sec:ParticleProperties}) for different configurations of its diffusion matrix (see upper left corner of left plots). The state of the net densities is shown at different evolution times (various colored and dashed lines) beginning at the initial proper time $\tau_0 = 2 \,\text{fm/c}$ (black, solid curve) and plotted over the spacetime rapidity $\eta_s$. The system is prepared at initial, homogeneous temperature $T_0 = 160 \, \text{MeV}$, a double-gaussian profile in net baryon number with maxima at $\eta_s = \pm 1.0$ with value $n_{B,\mathrm{max}} = 0.5 \, \mathrm{fm^{-3}}$ and initial vanishing local net strangeness density $n_{\mathrm{S}}$. }
			\label{fig:final_hydro_plot}
		\end{figure}

	We show our results for the evolution of the system for each of the four assumed cases in Fig.\ \ref{fig:final_hydro_plot} for proper times starting at $\tau = \tau_0 = 2$ fm/c until $\tau = 7$ fm/c. The evolution of the net baryon number (left side of the plot) and the net strangeness density (right side of the plot) is presented for the four cases: no diffusion (top row), full diffusion matrix (second row), no off-diagonal entries (third row) and baryon diffusion only (bottom row). Various colored and dashed curves are plotted over spacetime rapidity, $\eta_s$, representing the state at four different proper times: initial state at $\tau = 2$ fm/c (black solid curve), at $\tau = 3$ fm/c (blue dashed line), at $\tau = 5$ fm/c (orange dotted line), and finally at $\tau = 7$ fm/c (red mixed dashed line). We emphasize that two plots in one row belong to the same case. We show the densities multiplied by the Bjorken factor $\tau/\tau_0$ in order to account for the longitudinal expansion \cite{Bjorken1983}. 
	
	First, we note that in the non-diffusive case (Case 1), only a small amount of the baryon number is transported towards the mid- and outwards rapidities from the regions of high baryon densities due to the motion of the fluid (convection generated by pressure gradients). Furthermore, there is no transportation of net strangeness. Accordingly, in Case 2, there is also no extra transported net strangeness and the distribution of net strangeness remains flat at zero. However, in contrast to the non-diffusive case, there is significant diffusive transport of the net baryon number. All diffusive cases (Cases 2 to 4) show a very similar evolution of the net baryon number, but the evolution of the net strangeness is sensitive to the assumed configuration of the diffusion coefficient matrix. Contrary to Case 2, a wave-like profile in net strangeness density builds up over time in Cases 3 and 4, while the total net strangeness is conserved globally. This profile is more pronounced if the off-diagonal entry is neglected. To assess the magnitude of this effect, we compare the net strangeness density to the total number density, $n_{\mathrm{tot}}$. For Case 3, we find ratios up to $|n_{\mathrm{S}}/n_{\mathrm{tot}}| \sim 6\%$, and in the consideration of the full diffusion matrix (Case 4), only ratios up to $\sim 3\%$ are reached during the evolution in this example. Aside from the differences in magnitude, there are also differences in the wave-like profile that appears, depending on the assumed case. \newline
	
	\paragraph{Interpretation} \quad \\
	
	The reason for this separation of strangeness is that the Navier-Stokes terms of the corresponding diffusion currents introduce a coupling between the charge currents via the diffusion coefficient matrix (see Eq.~\eqref{eq:NavierStokesTerms}) \emph{and} the assumed equation of state.
		
	In Case 3, positive baryon number and positive strangeness is transported to the mid- and outward rapidity region. Due to charge conservation, less baryon number and negative net strangeness stays behind in the regions of the baryon source. From this case we see that even though we did not assume any explicit coupling through $\kappa_{\mathrm{SB}}$ in the fluid dynamic equations, we can still witness a correlation between the conserved charges. The origin of this intrinsic correlation of charges introduced by the equation of state \emph{alone} is the same as the correlation introduced by the diffusion matrix: the particles carry a multitude of conserved charge types. This in turn results in the fact that chemical potentials are generally dependent on all assumed charge densities and vice versa. Thus, this chemistry of "mixed" charges already encodes charge-correlation into the equation of state, see e.g.\ \cite{Monnai:2019hkn}. However, in order to achieve physically correct results for the charge-correlation during the dynamic evolution, it is important to ensure that the same chemistry is assumed for the calculation of the diffusion matrix as well. 
	
	This is demonstrated with Case 4, where we assumed the full diffusion coefficient matrix. We find a similar picture as in Case 3. However, because $\kappa_{\mathrm{SB}}$ is negative (as shown in Section \ref{sec:HadronicGasResults}), and the gradients in $\alpha_q$ have the same sign, the influences of both gradients on the diffusion currents inhibit or even cancel each other out in this configuration, which leads to a different evolution of the net strangeness in comparison to Case 3. The exact effects of this off-diagonal entry in the diffusion matrix clearly depend on the profiles of temperature and chemical potentials in a complicated manner, since the diffusion coefficients are also a function of these quantities. \newline
	
	\paragraph{Summary} \quad \\
	
	In this section, we presented first results which imply that choosing an equation of state and diffusion coefficients in an inconsistent way could make a difference in the evolution of a system that consists of particles carrying a multitude of conserved quantum numbers. However, these investigations were done in a simple manner and results from the evolution were not transferred to particle spectra. The question of whether the influence of the full diffusion coefficient matrix is significant is therefore left for more sophisticated and detailed future works.

	\section{Conclusion and Outlook}
	\label{sec:Conclusion}
	In the first part of this paper, we introduced the diffusion coefficient matrix in order to account for the fact that, especially in hadronic gases, particles generally carry multiple types of conserved quantum numbers. We find that the mixed chemistry results in a coupling of all diffusion currents that correspond to the conservation of these conserved quantities. In order to describe heavy ion collision, we propose that this coupling must be accounted for in dynamic simulations. 
	We evaluated the complete diffusion coefficient matrix for two examples: a hadron gas and a simple model for a massless QGP, both containing conserved baryon number, strangeness and electric charge. This was done by using a semi-analytical linear response approach in relativistic kinetic theory, and we compared our results for the diagonal coefficients to Ref.~\cite{Rougemont:2015ona} in the case of the massless, conformal QGP. We find that the off-diagonal coefficients $\kappa_{\mathrm{BQ}}$, $\kappa_{\mathrm{SB}}$ and $\kappa_{\mathrm{SQ}}$, which describe the mixing between the diffusion currents, can reach similar magnitudes to the diagonal coefficients, $\kappa_{\mathrm{BB}}$, $\kappa_{\mathrm{QQ}}$ and $\kappa_{\mathrm{SS}}$, which are usually evaluated in other approaches e.g. Ref.~\cite{Rougemont:2015ona,Soloveva:2019xph}. 
	
	Dynamic simulations or other model descriptions of high density heavy ion collisions in experiments like RHIC BES, NICA or FAIR will become increasingly important. We used the evaluated diffusion coefficient matrix and presented a first study of the influence of the matrix in a simple (1+1)D-fluid dynamic simulation of a hadronic system. In addition, signals of strangeness separation and significant baryon diffusion were found and discussed. The results imply that inconsistently choosing the equation of state and the diffusion coefficient matrix of the system results in false dynamics of the conserved charge, which could mislead the physical interpretation. We therefore advise that the mixing between the diffusion currents should not be neglected in simulations of high density heavy ion collisions. However, the relevance of these effects for experimental observables has not yet been investigated. Furthermore, significant effects from e.g.\ transverse dynamics, shear viscosity and second order contributions to the diffusion currents are expected, but were neglected in this first investigation. This, as well as other aspects remain open for more sophisticated works in the future. Moreover, a comparison of our results to lattice QCD, other transport models or dynamic approaches are also desirable for future research.

	\newpage
	
	\section{Appendix}
	
	\subsection{Particle properties of the hadrons and cross sections}
	\label{sec:ParticleProperties}
	
	\begin{table}[h!]
		\begin{tabular}{|c|c|c|c|c|c|c|}
			\hline
			Name & Mass [MeV/$c^2$] & Spin & Degeneracy & Baryon Number & Electric Charge & Strangeness \\
			\hline
			$\pi^+$ 		&	$138$	&	$0$		&	$1$	&	$0$		&	$+e$	&	$0$ 	\\
			$\pi^-$ 		&	$138$	&	$0$		&	$1$	&	$0$		&	$-e$	&	$0$ 	\\
			$\pi^0$ 		&	$138$	&	$0$		&	$1$	&	$0$		&	$0$		&	$0$ 	\\
			$K^+$ 			&	$496$	&	$0$		&	$1$	&	$0$		&	$+e$	&	$+1$ 	\\
			$K^-$ 			&	$496$	&	$0$		&	$1$	&	$0$		&	$-e$	&	$-1$ 	\\
			$K^0$ 			&	$496$	&	$0$		&	$1$	&	$0$		&	$0$		&	$+1$ 	\\
			$\bar{K}^0$ 	&	$496$	&	$0$		&	$1$	&	$0$		&	$0$		&	$-1$ 	\\
			$p$ 			&	$938$	&	$1/2$	&	$2$	&	$+1$	&	$+e$	&	$0$ 	\\
			$\bar{p}$ 		&	$938$	&	$1/2$	&	$2$	&	$-1$	&	$+e$	&	$0$ 	\\
			$n$ 			&	$938$	&	$1/2$	&	$2$	&	$+1$	&	$0$		&	$0$ 	\\
			$\bar{n}$ 		&	$938$	&	$1/2$	&	$2$	&	$-1$	&	$0$		&	$0$ 	\\
			$\Lambda^0$ 		&	$1116$	&	$1/2$	&	$2$		&	$+1$	&	$0$		&	$-1$ 	\\
			$\bar{\Lambda}^0$ 	&	$1116$	&	$1/2$	&	$2$		&	$-1$	&	$0$		&	$+1$ 	\\
			$\Sigma^0$ 			&	$1193$	&	$1/2$	&	$2$		&	$+1$	&	$0$		&	$-1$ 	\\
			$\bar{\Sigma}^0$ 	&	$1193$	&	$1/2$	&	$2$		&	$-1$	&	$0$		&	$+1$ 	\\
			$\Sigma^+$ 			&	$1189$	&	$1/2$	&	$2$		&	$+1$	&	$+e$	&	$-1$ 	\\
			$\bar{\Sigma}^+$ 	&	$1189$	&	$1/2$	&	$2$		&	$-1$	&	$-e$	&	$+1$ 	\\
			$\Sigma^-$ 			&	$1197$	&	$1/2$	&	$2$		&	$+1$	&	$-e$	&	$-1$ 	\\
			$\bar{\Sigma}^-$ 	&	$1197$	&	$1/2$	&	$2$		&	$-1$	&	$+e$	&	$+1$ 	\\
			\hline
		\end{tabular}
		\caption{Properties of the particle species used in the hadronic calculation of the diffusion coefficient matrix. Here, $e$ denotes the elementary electric charge.}
	\end{table}
	
	\begin{table}[h!]
		\begin{tabular}{|c||c|c|c|c|c|c|c|c|c|c|c|c|c|c|c|c|c|c|c|}
			\hline
			& $\pi^+$ 	&$\pi^-$& $\pi^0$	&$K^+$ 		& $K^-$ & $K^0$ & $\bar{K}^0$ & $p$ & $n$ & $\bar{p}$ & $\bar{n}$ & $\Lambda^0$ & $\bar{\Lambda}^0$ & $\Sigma^0$ & $\bar{\Sigma}^0$ & $\Sigma^+$ & $\bar{\Sigma}^+$ & $\Sigma^-$ & $\bar{\Sigma}^-$ \\
			\hhline{|=#=|=|=|=|=|=|=|=|=|=|=|=|=|=|=|=|=|=|=|}
			$\pi^+$ 	& 10 	&  	\res		& 	\res		& 10	& 10 & \res & 10 & \res & 10 & 10 & \res & 23.1 & 23.1 & 5 & 5 & 5 & 5 & 5 & 5 \\
			\hline
			$\pi^-$ 	&  		& 10		& 	\res		& \res  & 10 & 10 & \res  & \res & \res & \res & 10 & 23.1 & 23.1 & 5 & 5 & 5 & 5 & 5 & 5 \\
			\hline
			$\pi^0$	 	&  		&  			& 5 		& \res  & 10 & \res  & \res  & \res & \res & \res &  \res & 23.1 & 23.1 & 5 & 5 & 5 & 5 & 5 & 5 \\
			\hline
			$K^+$ 		&  		&  			& 		 	& 10 	& 10 	& 10 	& 50 	&  \res	& 10 	& 20 	& 10 & 18.5 & 18.5 & 3 & 3 & 3 & 3 & 3 & 3 	\\  	    
			\hline	
			$K^-$ 		&  		&  			&  			&  		& 10 	& 50 	& 10 	&  \res & \res  	& 6 	& 10 & 18.5 & 18.5 & 3 & 3 & 3 & 3 & 3 & 3 	\\ 			
			\hline
			$K^0$ 		&  		&  			&  			&  		&  		& 10 	& 50 	& 6 & 6 	& 20 	& 20 & 18.5 & 18.5 & 3 & 3 & 3 & 3 & 3 & 3	\\			
			\hline
			$\bar{K}^0$ &  		&  			&  			&  		&  		&  		& 10 	& 8 & 20 	& 6 	& 6 & 18.5 & 18.5 & 3 & 3 & 3 & 3 & 3 & 3	\\			
			\hline
			$p$ 		&  		&  			&  			& 	 	&  		&  		&  		& \res 	&  	\res	&  \res		& 20 & 34.7 & 34.7 & 10 & 10 & 10 & 10 & 10 & 10	\\			
			\hline
			$n$ 		&  		&  			&  			&  		&  		&  		&  		&  	& 	 20 & 	\res	& 100 & 34.7 & 34.7 & 10 & 10 & 10 & 10 & 10 & 10 	\\			
			\hline
			$\bar{p}$ 	&  		& 			&  &  &  &  &  &  &  & 10 & 10  & 34.7 & 34.7 & 10 & 10 & 10 & 10 & 10 & 10	\\	
			\hline
			$\bar{n}$ 	&  		&  			&  &  &  &  &  &  &  & & 10 & 34.7 & 34.7 & 10 & 10 & 10 & 10 & 10 & 10	\\	
			\hline
			$\Lambda^0$ 	&  		&  			&  &  &  &  &  &  &  &  &  & 30 & 30 & 10 & 10 & 10 & 10 & 10 & 10	\\	
			\hline
			$\bar{\Lambda}^0$ 	&  		&  			&  &  &  &  &  &  &  &  &  &  & 30 & 10 & 10 & 10 & 10 & 10 & 10	\\	
			\hline
			$\Sigma^0$ 	&  		&  			&  &  &  &  &  &  &  &  &  &  &  & 10 & 10 & 10 & 10 & 10 & 10	\\	
			\hline
			$\bar{\Sigma}^0$ 	&  		&  			&  &  &  &  &  &  &  &  &  &  &  &  & 10 & 10 & 10 & 10 & 10	\\	
			\hline
			$\Sigma^+$ 	&  		&  			&  &  &  &  &  &  &  &  &  &  &  &  &  & 10 & 10 & 10 & 10	\\	
			\hline
			$\bar{\Sigma}^+$ 	&  		&  			&  &  &  &  &  &  &  &  &  &  &  &  &  &  & 10 & 10 & 10	\\	
			\hline
			$\Sigma^-$ 	&  		&  			&  &  &  &  &  &  &  &  &  &  &  &  &  &  &  & 10 & 10	\\	
			\hline
			$\bar{\Sigma}^-$ 	&  		&  			&  &  &  &  &  &  &  &  &  &  &  &  &  &  &  &  & 10	\\	
			\hline
		\end{tabular}
		\caption{All elastic cross sections among all species. The constant cross sections are in units of mb, the label \res ~refers to the tabulated or parametrized resonance cross sections depicted in Fig.~\ref{fig:all_resonances_paper}. We use constant cross sections where no resonance cross section was available from UrQMD \cite{UrQMD1,UrQMD2}}.
		\label{tab:1}
	\end{table}
	\newpage

	\subsection{Fluid dynamic equations: (1+1)-dimensional longitudinal system in hyperbolic coordinates without bulk and shear viscosity}
	\label{sec:ExplicitFluidDynamics}
	
	The transformation law between cartesian and hyperbolic coordinates $(t,x,y,z) \leftrightarrow (\tau, x, y, \eta_s)$ reads
	\begin{align}
		t = \tau \cosh(\eta_s), \quad x = x, \quad y = y, \quad z = \tau \sinh(\eta_s),
	\end{align}
	where $\tau$ is the proper time and $\eta_s$ is the spacetime rapidity. The metric in hyperbolic coordinates reads $g_{\mu\nu} = \mathrm{diag}\left(1, -1, -1, -\tau^2 \right)$ and the only non-vanishing Christoffel symbols of second kind are: $\Gamma^\tau_{\eta\eta} = \tau$ and $\Gamma^\eta_{\tau\eta} = \Gamma^\eta_{\eta\tau} = \frac{1}{\tau}$. The fluid velocity simplifies to $u^\mu = \gamma_\eta \left(1, 0, 0, v^\eta \right)$, with the Lorentz factor $\gamma_\eta = \frac{1}{\sqrt{1 - \tau^2 (v^\eta)^2}}$, and the four-derivative reading $\partial_\mu = \left(\partial_\tau, 0, 0, \partial_\eta \right)$. 
	The expansion scalar can then be expressed as
	\begin{align}
		\theta \equiv \nabla_\mu u^\mu = \partial_\tau \gamma_\eta + \partial_\eta \left( \gamma_\eta v^\eta \right) + \frac{\gamma_\eta}{\tau} .
	\end{align}
	We set all initial values of the dissipative fields to zero. Therefore, the non-vanishing fluid dynamic fields read:
	\begin{align}
		T^{\tau\tau} &= \left(\epsilon + P_0 \right) \gamma_\eta^2 - P_0 \\
		T^{\tau\eta} &= (T^{\tau\tau} + P_0)v^\eta \\
		T^{ii} &= P_0, \quad i \in \lbrace x,y \rbrace \\
		T^{\eta\eta} &= T^{\tau\eta} v^\eta + \frac{P_0}{\tau^2} \\
		\nonumber \\
		N^\tau_{q} &= n_{q} \gamma_\eta + j^\tau_{q} \\
		N^\eta_{q} &= \left(N^\tau_{q} - j^\tau_{q} \right) v^\eta + j^\eta_{q}.
	\end{align}
	Due to the orthogonality of the diffusion current, $j^\eta_{q}$ is the only independent component, and therefore
	\begin{align}
		j^\tau_{q} = \tau^2 v^\eta j^\eta_{q}.
	\end{align}
	Using the explicit form of the fluid dynamic tensor, we can express local rest frame quantities in terms of the lab frame quantities as
	\begin{align}
		\epsilon &= T^{\tau\tau} - T^{\tau\eta} \tau^2 v^\eta, \label{eq:energdens}\\
		n_{q} &= \frac{1}{\gamma_\eta} \left(N^\tau_{q} - j^\tau_{q} \right). \label{eq:chargedens}
	\end{align} 
	Further, the fluid velocity can also be connected to the lab frame quantities,
	\begin{align}
		v^\eta = \frac{T^{\tau\eta}}{T^{\tau\tau} + P_0}, \label{eq:velocityeta}
	\end{align}
	and is evaluated using Newtons secant algorithm because the components of the energy-momentum tensor are dependent on the fluid velocity itself. 
	
	In order to evaluate the primary fluid dynamic fields - $\epsilon$, $n_{q}$ and $v^\eta$ - the lab frame quantities - $T^{\tau\tau}$, $T^{\tau\eta}$, $N^\tau_{q}$ and $j^\eta_{q}$ - must be calculated by solving the fluid dynamic equations of motion. The explicit form of these read:
	\begin{align}
		\partial_\tau T^{\tau\tau} + \partial_\eta \left( v^\eta T^{\tau\tau} \right) &= -\partial_\eta \left( v^\eta P_0 \right) - \frac{1}{\tau} T^{\tau\tau} - \tau T^{\eta\eta}, \label{eq:Ttautau} \\
		\partial_\tau T^{\tau\eta} + \partial_\eta\left( v^\eta T^{\tau\eta} \right) &= -\frac{1}{\tau^2}\partial_\eta P_0 - \frac{3}{\tau} T^{\tau\eta}, \label{eq:Ttaueta} \\
		\partial_\eta N^\tau_{q} + \partial_\eta \left( v^\eta N^\tau_{q} \right) &= -\frac{N^\tau_{q}}{\tau} + \partial_\eta\left( v^\eta j^\eta_{q} \right) - \partial_\eta j^\eta_{q}, \label{eq:Ntau}
	\end{align}
	and the equation of motion for the diffusion current in $\eta_s$-direction reads:
	\begin{align}
		\left( \partial_\tau  + v^\eta \partial_\eta \right) j^\eta_{q} &= -\sum_{q^\prime}  \frac{\kappa_{qq^\prime}}{\tau_{q} \gamma_\eta}\left[ \frac{1}{\tau^2} \partial_\eta \alpha_{q^\prime} + \gamma_\eta^2 v^\eta \left( \partial_\tau + v^\eta \partial_\eta \right) \alpha_{q^\prime} \right] \nonumber \\ 
		& \quad \quad \quad \quad \quad \quad - \frac{j^\eta_{q}}{\tau_{q} \gamma_\eta} - \frac{1}{\tau}\left( j^\eta_q + v^\eta j^\tau_q \right) - v^\eta \left( j^\tau_{q} \mathcal{D}u_\tau + j^\eta_{q} \mathcal{D}u_\eta \right), \label{eq:DiffEta}
	\end{align}
	and 
	\begin{align}
		\mathcal{D}u_\tau &=  \gamma_\eta \left( \partial_\tau \gamma_\eta + v^\eta \partial_\eta \gamma_\eta \right) + \tau \gamma_\eta^2 (v^\eta)^2 , \\
		\mathcal{D}u_\eta &= -\tau^2 \gamma_\eta \left[ \partial_\tau \left( \gamma_\eta v^\eta \right) + v^\eta \partial_\eta \left( \gamma_\eta v^\eta \right) \right] - 2\tau \gamma_\eta^2 v^\eta.
	\end{align}
	Further, we propose simple estimates for the relaxation times motivated from Ref.~\cite{Denicol2012} 
	\begin{align}
		\tau_q \equiv \frac{12 \kappa_{qq}}{n_{\mathrm{tot}}} .
	\end{align}
	Continuity equations of the form 
	\begin{align}
		\partial_\tau \rho(\tau,\eta) + \partial_\eta \left[ v^\eta(\tau,\eta) \rho(\tau,\eta) \right] = \mathcal{S}(\tau,\eta),
	\end{align}
	where $\rho$ is the evolving quantity and $\mathcal{S}$ is a source term, can be solved by applying the numerical solving scheme SHASTA \cite{BorisBook}. We do so in the same manner as is done in Ref.~\cite{Molnar2009a}.
	
	\section*{Acknowledgments}
	The authors M.G., J.A.F. and C.G. acknowledge support from the Deutsche
	Forschungsgemeinschaft (DFG) through the grant CRC-TR 211 \textquotedblleft
	Strong-interaction matter under extreme conditions\textquotedblright . M.G. and J.A.F.
	acknowledge support from the \textquotedblleft Helmholtz Graduate School
	for Heavy Ion research\textquotedblright. J.A.F. acknowledges support from the  \textquotedblleft Stiftung Polytechnische
	Gesellschaft\textquotedblright, Frankfurt am Main.  Furthermore, G.S.D. would like to thank Conselho Nacional de
	Desenvolvimento Cient\'{\i}fico e Tecnol\'{o}gico (CNPq) and Funda\c c\~ao de Amparo \`a Pesquisa do Estado do Rio de Janeiro (FAPERJ) for financial support. H.N. is supported by the Academy of Finland, Project no. 297058.
	\bibliographystyle{apsrev4-1}
	\bibliography{paper_prd.bib} 
	
\end{document}